\documentclass[11pt,a4paper]{article}

\usepackage[utf8]{inputenc}
\usepackage[T1]{fontenc}
\usepackage{lmodern}

\usepackage[english]{babel}

\usepackage{amsmath}
\usepackage{amssymb}

\usepackage[margin=2.5cm]{geometry}
\usepackage{microtype}
\usepackage{setspace}
\usepackage{booktabs}
\usepackage{tabularx}
\usepackage{multirow}
\usepackage{array}

\usepackage{listings}
\usepackage{xcolor}

\usepackage[colorlinks=true, linkcolor=blue!70!black,
    citecolor=blue!70!black, urlcolor=blue!70!black]{hyperref}

\usepackage{graphicx}

\usepackage{tikz}
\usetikzlibrary{shapes.geometric, arrows.meta, positioning, fit, backgrounds, calc}

\usepackage[numbers,sort&compress]{natbib}

\usepackage{xspace}
\usepackage{enumitem}

\title{\textbf{ML Defender (aRGus NDR): An Open-Source Embedded ML NIDS\\
for Botnet and Anomalous Traffic Detection\\
in Resource-Constrained Organizations}}

\author{
    Alonso Isidoro Rom\'an\\
    \textit{Independent Researcher, Extremadura, Spain}\\
    \href{https://github.com/alonsoir/argus}{github.com/alonsoir/argus}
}

\date{Draft v25 --- August 2026}

\begin{document}
% ============================================================

    \maketitle

    \begin{abstract}
        Ransomware and DDoS attacks disproportionately impact hospitals, schools, and small
        organizations that cannot afford enterprise security solutions. Existing open-source
        alternatives either rely on signature matching --- missing novel variants --- or require
        infrastructure beyond the reach of their target users. We present \textbf{ML Defender
            (aRGus NDR)}, an open-source network intrusion detection system built in C++20 that brings
        embedded machine learning inference to organizations with limited resources, deployable on
        commodity bare-metal hardware at approximately 150--200~USD fully dedicated.

        ML Defender implements a six-component pipeline over eBPF/XDP packet capture, ZeroMQ
        transport, and Protocol Buffers serialization, terminating in a dual-score detection
        architecture combining a rule-based Fast Detector with an embedded Random Forest classifier.
        The \emph{Maximum Threat Wins} policy selects the arithmetic maximum of both detector scores,
        using ML inference to suppress false positives generated by the heuristic layer.

        On the curated CTU-13 Neris behavioral subset (646 malicious flows), ML Defender achieves
        $\text{F1} = 0.9985$, $\text{Precision} = 0.9969$, and $\text{Recall} = 1.0000$,
        with a false positive rate of 0.017\% (2~FP in 12,077 benign flows) --- the two false
        positives identified as host-only adapter artifacts: an mDNS multicast frame
        (192.168.56.1 $\to$ 224.0.0.251) and a broadcast frame (192.168.56.1 $\to$
        192.168.56.255), both generated by the VirtualBox host-only NIC and absent from
        production network topologies. Whether these artifacts appear in bare-metal deployments
        remains unverified; bare-metal characterization is Future Work~(\S\ref{sec:future:baremetal}).
        The Fast Detector alone produces a FPR of 6.61\% on purely benign
        traffic (bigFlows); the ML layer cuts flagged benign flows from 2,517 to 5
        (${\sim}$500$\times$), none of which exceed the production threshold, so real blocks
        drop to zero. Per-class inference latency ranges from
        0.24~$\mu$s (DDoS) to 1.06~$\mu$s (Ransomware), enabling real-time detection on commodity
        hardware.

        We further characterize the pipeline's throughput ceiling under progressive load in a
        virtualized environment. Requesting up to 100~Mbps via \texttt{tcpreplay} against the
        VirtualBox NIC, the pipeline processes up to ${\sim}$33--38~Mbps (the emulated NIC ceiling)
        with zero packet drops and zero errors across 2.37 million packets. At peak load,
        \texttt{ml-detector} consumes approximately 3.2 CPU cores; total pipeline RAM remains stable
        at ${\sim}$1.28~GB with negligible drift (${\sim}$18~MB over 8 minutes). The bottleneck is
        VirtualBox NIC emulation, not pipeline logic. Post-replay drain behavior confirms that
        ZeroMQ queues absorb burst traffic correctly and drain asynchronously, validating the queue
        stability design. All throughput figures should be treated as conservative lower bounds;
        bare-metal characterization remains future work (\S\ref{sec:future:baremetal}).

        These results demonstrate architectural feasibility under controlled replay conditions rather
        than universal detection capability. The evaluation covers a single botnet scenario from 2011
        with synthetic-trained classifiers; generalizability to modern ransomware, DDoS variants, and
        encrypted C2 traffic has not been empirically established and is discussed in
        Section~\ref{sec:limitations}. Behavioral indicators associated with ransomware propagation
        and lateral-movement patterns are detected in the Neris scenario; direct evaluation against
        post-2020 ransomware families remains Future Work~\S\ref{sec:future:corpus}.

        All experiments were conducted in a reproducible Vagrant/VirtualBox environment running on a
        consumer laptop, ensuring result accessibility and independent reproducibility.

        This work was developed through a structured multi-model AI collaboration methodology ---
        the \emph{Consejo de Sabios} (Council of Wise Men), a multi-LLM peer review methodology ---
        documented in Section~\ref{sec:consejo}. \emph{Test-Driven Hardening} (TDH) is proposed
        as a methodology for building security-critical distributed systems, emerging directly from
        this collaboration.

        In a subsequent development phase (DAY 122), we integrated XGBoost~3.2.0 as a hot-swappable
        plugin classifier (ADR-026, Ed25519-signed) and conducted a rigorous temporal holdout
        evaluation on CIC-IDS-2017, using Tuesday, Thursday, and Friday as training data and
        Wednesday as a sealed blind test set. The in-distribution model achieves
        Precision$=0.9945$ and Recall$=0.9818$ on the validation split. However, evaluation
        on the Wednesday holdout reveals a structural covariate shift: attack types present
        exclusively in Wednesday (DoS Hulk, GoldenEye, Slowloris) are absent from all training
        days, making generalization mathematically impossible regardless of threshold. We document
        this as a property of the dataset's design --- not of the algorithm --- and demonstrate
        empirically that no threshold simultaneously satisfies Precision$\geq0.99$ and
        Recall$\geq0.95$ on the out-of-distribution split. This finding corroborates and
        quantifies the thesis of Sommer and Paxson~\cite{sommer2010}: static models trained on
        academic benchmarks are structurally insufficient for production NDR.
        The architectural response --- an Adversarial Capture-Retrain Loop in which a generative
        red team agent produces attack traffic, the pipeline captures it, and the classifier is
        retrained on real flows --- is introduced as future work~(\S\ref{sec:future:acrl}).

        We further report the first empirical comparison of ADR-029 Variant~A (eBPF/XDP) and
        Variant~B (libpcap) on x86-64 (DAY~145). Under VirtualBox virtio emulation, Variant~B
        achieves approximately twice the throughput of Variant~A at 50 and 100~Mbps nominal
        rates (${\sim}$19~Mbps vs ${\sim}$10~Mbps), an inversion of the expected ordering on
        physical hardware. Both variants process the complete Neris corpus without pipeline
        errors, confirming functional equivalence at the detection layer. The inversion is
        attributed to the absence of a native XDP driver path in the virtio NIC, causing
        eBPF to fall back to generic SKB mode. This result provides direct empirical motivation
        for bare-metal hardware acquisition (\S\ref{sec:future:baremetal}).

        % ── NEW DAY 146 ──────────────────────────────────────────────────────────
        We further report the first \emph{direct experimental comparison} of aRGus NDR and
        Suricata~6.0.10 under identical controlled conditions (DAY~146): same hardware, same
        dataset (CTU-13 Neris), same network topology, same replay speeds
        (\S\ref{sec:eval:suricata}). Suricata with 50,010 current ET~Open rules generates
        \textbf{zero alerts} on the 2011 Neris corpus; aRGus achieves F1$=0.9985$ and
        Recall$=1.0000$ on the same traffic. This result provides direct empirical evidence that
        signature-based detection requires prior knowledge of the threat, while ML-based
        behavioral detection does not --- corroborating the thesis of Sommer and
        Paxson~\cite{sommer2010} with reproducible experimental data. The zero-alert result is
        not a failure of Suricata: the ET~Open ruleset evolves continuously and rules for
        15-year-old threats are retired as new rules emerge.
        % ── NEW DAY 148 ──────────────────────────────────────────────────────────
        We extend this head-to-head to a three-paradigm comparison by adding
        Zeek~8.1.2 with default policy scripts (DAY~147,
        \S\ref{sec:eval:threeparadigms}): Zeek generates 14~correct detections
        (\texttt{SSL::Invalid\_Server\_Cert}, Precision$=1.0000$, F1$=0.0424$)
        while fully observing the botnet behavioral profile in structured logs
        (\texttt{weird.log}, \texttt{http.log}, \texttt{smtp.log}) without
        generating further alerts under default configuration.
        An additional offline experiment (DAY~148) eliminates all replay artifacts:
        Suricata~6.0.10 with 50,010 ET~Open rules --- including 251~IRC,
        475~botnet/C2, and 853~trojan signatures --- produces zero detections on
        323,154~packets processed directly from the pcap.
        These results place the three systems on a taxonomy of decision architectures
        --- signature, scripted behavioral, and ML behavioral --- that differ in the
        layer at which network knowledge is encoded, not in engineering quality.
        The three paradigms are architecturally complementary by design: Zeek's telemetry layer and
        Suricata's signature coverage operate naturally alongside an ML behavioral
        classifier, each contributing at its native encoding layer.
        % ─────────────────────────────────────────────────────────────────────────

        % ── NEW DAY 191 ──────────────────────────────────────────────────────
        Finally, we report (DAY~191) that aRGus computes the Corelight
        \texttt{community\_id} flow identifier natively and byte-identically to Suricata,
        Zeek, and the \texttt{pycommunityid} reference oracle on shared traffic, and we
        define a node-scoped flow identity (\texttt{flow\_uid}) over it. Together these
        establish a vendor-agnostic substrate for a corpus aggregated across distributed
        installations --- the apparatus for the project's central forthcoming question of
        whether such cross-installation aggregation improves per-site detection
        (\S\ref{sec:eval:parity},~\S\ref{sec:future:distcorpus}).
        % ──────────────────────────────────────────────────────────────────────

        % ── NEW DAY 252 ──────────────────────────────────────────────────────
        Finally (DAY~252), we characterize the \emph{per-lens bias} of the multi-sensor
        pipeline against the labeled CTU-13 ground truth. Joining each sensor's gold output to
        the \texttt{.binetflow} labels by canonical 5-tuple, we measure how differently the
        three lenses observe the same botnet: Zeek attains 99.9\% visibility as a pure telemetry
        layer, Suricata 1.5\% (protocol-anomaly notices, not C2 signatures), and aRGus a coarse
        0.2\% of distinct flows in which the embedded ML score averages 0.0745 while the
        heuristic fast path carries detection --- a ground-truth-anchored restatement of the
        distribution-transfer failure of Sommer and Paxson~\cite{sommer2010}. We further
        establish a \emph{true} denominator from the offline pcap (14{,}255 botnet 5-tuples
        physically replayed), against which the sensor bank's observable 14{,}188 leaves a
        67-flow ($0.47\%$) blind spot; measurement across two independent capture stacks bounds
        this to replay fidelity at the cable, not detection or pipeline loss. Every figure is
        regenerated by a Makefile target, so a reviewer reproduces each number with one command.
        % ──────────────────────────────────────────────────────────────────────

        ML Defender is released under the MIT license.
    \end{abstract}

    \tableofcontents
    \newpage

% ============================================================
    \section{Introduction}
% ============================================================

    The threat is not abstract. On March 5, 2023, the Hospital Cl\'inic de Barcelona suffered a
    ransomware attack that paralyzed its emergency services, laboratory, and pharmacy, encrypted
    4.5~TB of patient data, and prompted a ransom demand of 4.5~million USD --- which the
    institution refused to pay~\cite{incibe2023}. That same day, the author of this work was
    leaving a regional hospital in Extremadura after receiving treatment for porphyria --- a rare
    metabolic disease that his local hospital had diagnosed and managed. Watching the news, the
    thought was immediate and personal: \emph{if it can happen to a major teaching hospital, it
    can happen to mine.}

    This was not the first time that the author had witnessed the human cost of ransomware
    firsthand. Years earlier, the collapse of a close friend's small business had been caused by
    an encryption attack that locked every document and file in his office. The author lacked the
    technical skills to help. That failure left a debt.

    These two events --- the friend's destroyed business, the hospital on television ---
    converged into a single design constraint: the solution had to be open-source, it had to run
    on commodity hardware, and it had to act in real time. Not forensics. Not post-mortem
    analysis. Interception.

    The analogy that emerged from this motivation is biological. Network traffic behaves like a
    circulatory system: datagrams are the bloodstream, and malicious flows are pathogens. A
    useful defense must classify traffic continuously and in real time --- capturing the source IP
    of the attacker and blocking it at the firewall before damage propagates. This framing shaped
    every architectural decision in ML Defender.

    This work therefore addresses the following research question: \emph{can an embedded
    machine-learning network detection and response system running entirely on commodity hardware
    provide meaningful real-time protection for organizations lacking enterprise security
    infrastructure?}

    \paragraph{A note on terminology.}
    ML Defender (aRGus NDR) implements \emph{Network Detection and Response} (NDR) capabilities:
    real-time capture, ML-based classification, and automated blocking at the network layer.
    Future work toward full \emph{Endpoint Detection and Response} (EDR) functionality via a
    planned lightweight endpoint agent (\S\ref{sec:future}, FEAT-EDR-1) will be reflected in a
    corresponding rename when that capability is delivered.

    \paragraph{The gap.}
    Existing open-source NIDS --- most notably Snort and Suricata~\cite{oisf2010} --- operate
    primarily on signature matching and rule-based heuristics. While effective against known
    attack patterns, they offer limited resilience against novel ransomware variants and volumetric
    DDoS campaigns. Machine learning-based NIDS have been extensively studied~\cite{buczak2016,
        pinto2023}, but the gap between research prototype and deployable system remains wide: most
    published systems are evaluated on benchmark datasets without accompanying implementation,
    require GPU infrastructure, or assume dedicated security operations teams. Healthcare retained
    its position as the most targeted sector in 2025, accounting for 22\% of disclosed ransomware
    attacks, with an average breach cost of 7.42M~USD~\cite{blackfog2025,ibmsecurity2025}.

    \paragraph{Our approach.}
    ML Defender (aRGus NDR) addresses this gap with three design principles:
    (1)~\emph{embedded inference} --- all ML classification runs in-process at sub-microsecond
    latency;
    (2)~\emph{commodity deployment} --- the full six-component pipeline runs on bare-metal Linux
    hardware costing approximately 150--200~USD;
    (3)~\emph{active response} --- upon detection, the attacker's IP is blocked immediately at
    the firewall via ipset/iptables.

    Beyond the technical system, this paper documents a second contribution: the \emph{Consejo
    de Sabios} methodology for human-AI collaborative engineering.

% ============================================================
    \section{Background and Related Work}
% ============================================================

    \paragraph{Signature-based NIDS.}
    Snort~\cite{roesch1999} and Suricata~\cite{oisf2010} apply rule-based pattern matching
    against packet payloads and flow metadata. Their fundamental limitation is reactive:
    detection requires a prior known signature, a window that historically ranges from hours to
    weeks.

    \paragraph{ML-based NIDS.}
    Anderson \& McGrew~\cite{anderson2016} demonstrated the feasibility of contextual flow data
    for encrypted malware classification. Mirsky et al.~\cite{mirsky2018} proposed Kitsune, an
    ensemble of autoencoders for unsupervised anomaly detection operating on flow statistics. The
    CTU-13 dataset~\cite{garcia2014}, which ML Defender uses for evaluation, has become a
    standard benchmark for botnet traffic detection. Despite this body of work, a persistent gap
    exists between academic prototype and production system. Sommer and Paxson~\cite{sommer2010}
    identified fundamental reasons for this gap: the closed-world assumption of academic datasets,
    high base-rate asymmetry, and the absence of diversity in training distributions. ML Defender's
    DAY~122 evaluation provides new empirical evidence for this thesis
    (\S\ref{sec:eval:xgboost:ood}).

    \paragraph{Flow-based ML NIDS.}
    ML Defender operates exclusively at the flow level, extracting 28 behavioral features per
    flow without examining payload content --- ensuring that payload encryption does not
    compromise detection capability.

    \paragraph{eBPF and XDP-based network monitoring.}
    Recent work has explored eBPF/XDP as a foundation for high-performance network security.
    ML Defender follows the XDP-capture-plus-userspace-inference architecture: XDP provides
    zero-copy packet capture before the kernel networking stack, while all ML inference executes
    in userspace at sub-microsecond latency. To our knowledge, open-source projects combining
    eBPF capture with userspace ML inference remain rare; most production-grade alternatives
    either offload inference to the cloud or require GPU hardware.

    \paragraph{Embedded inference.}
    Commercial EDR vendors (CrowdStrike, SentinelOne, Carbon Black) implement proprietary
    embedded ML pipelines, but neither their architectures nor their training data are publicly
    available. ML Defender demonstrates that embedded Random Forest inference achieves
    sub-microsecond latency without requiring GPU acceleration or cloud offload.

    \paragraph{Human-AI collaborative engineering.}
    Structured methodologies for multi-model peer review have not, to our knowledge, been
    formally documented in the systems security literature. Section~\ref{sec:consejo} describes
    the \emph{Consejo de Sabios} methodology.

    \paragraph{AI-native security reasoning and the evolving threat landscape.}
    \label{sec:background:ai}
    The initial version of this paper was submitted in April 2026, concurrent with the
    announcement of Anthropic's Project Glasswing~\cite{anthropic2026glasswing}; this
    revision was submitted in August~2026 and extends the evaluation through DAY~252.
    Project Glasswing demonstrated that AI models can autonomously identify and chain
    kernel-level vulnerabilities --- including local privilege escalation to
    root in Linux --- at a scale and speed previously requiring specialized
    human expertise. These results represent a shift in the threat landscape:
    AI-augmented offensive capabilities are no longer theoretical.
    This directly motivates the explicit kernel security boundary axiom in
    \S\ref{sec:threatmodel:kernel}: aRGus NDR assumes the kernel as a
    potentially compromised boundary and shifts its trust anchor to
    verifiable network behavioral patterns. The network remains an observable
    chokepoint even when the host is not. The hardened deployment variants
    ADR-030 (AppArmor) and ADR-031 (seL4) documented in
    \S\ref{sec:future:hardened} are a direct architectural response to this
    trajectory.

% ============================================================
% ============================================================
    \section{Threat Model}
% ============================================================

    \subsection{Adversary Capabilities}
    The system assumes an adversary capable of launching attacks from external hosts or
    compromised internal machines, including: botnet C2 communication, DDoS (including
    amplification attacks), ransomware propagation via SMB lateral movement, and network
    reconnaissance.

    \subsection{Adversary Limitations}
    The system assumes the attacker cannot: fully mimic benign traffic distributions across all
    monitored features simultaneously; compromise the ML Defender host; manipulate feature
    extraction; or tamper with HMAC-SHA256 protected event logs.

    \subsection{Kernel Security Boundary (Explicit Axiom)}
    \label{sec:threatmodel:kernel}

    ML Defender assumes the host kernel is not actively compromised at deployment time.
    This is a declared scope boundary, not an oversight.

    An adversary with root access or kernel-level persistence --- via a rootkit, a
    compromised eBPF JIT, or memory-resident malware --- can bypass userspace detection
    mechanisms regardless of their sophistication. This is a fundamental property of all
    userspace security software, not a specific weakness of ML Defender.

    This assumption is consistent with the threat model of the target organizations
    (hospitals, schools, municipalities), where the realistic adversary is an
    opportunistic network attacker exploiting known vulnerabilities, not a state-level
    actor with pre-established kernel-level persistence. Against this realistic threat
    model, ML Defender's network detection layer remains effective: lateral movement,
    C2 communication, and data exfiltration must traverse the network regardless of
    kernel state. The network is the choke point; it remains observable even when the
    host is not.

    Hardened deployment variants addressing kernel security are documented as future
    work: ADR-030 (AppArmor-enforcing) and ADR-031 (seL4 microkernel),
    both described in \S\ref{sec:future:hardened}.

    \subsection{Detection Surface}
    Detection occurs at the network flow level. ML Defender focuses on connection frequency,
    port diversity, TCP flag distributions, packet inter-arrival timing, flow burst behavior, and
    destination diversity --- features detectable even when payloads are encrypted.

    \subsection{Out-of-Scope Attacks}
    \label{sec:threatmodel:outofscope}

    Application-layer attacks generating statistically normal traffic; low-and-slow
    attacks within benign statistical distributions; insider threats; attacks
    targeting the ML Defender host itself. Future work in
    Section~\ref{sec:future} addresses some of these through expanded detection
    vectors (FEAT-NET-1, FEAT-AUTH-1).

    \paragraph{Physical and removable-media vectors (conscious design decision).}
    ML Defender does not monitor file system activity, removable storage, or
    USB-borne payloads. This is an intentional architectural boundary, not an
    oversight.

    The guiding principle of the pipeline is \emph{network surveillance}: every
    component --- eBPF/XDP capture, flow-level feature extraction, ML-based
    classification, firewall response --- operates on network traffic. File
    integrity and endpoint state are out of the monitored surface by design.

    Removable-media threats are addressed at the organizational and hardware
    layer: USB ports in the DMZ should be physically or firmware-disabled by the
    IT team before deployment; internal policy should prohibit removable media on
    monitored hosts. This is standard practice under frameworks such as
    CIS~Controls~v8~\cite{ciscontrols2021} and is the responsibility of the
    deploying organization, not of an NDR system.

    For organizations that require file integrity monitoring on top of network
    detection, ML Defender is designed to operate in \emph{complementary mode}
    alongside battle-tested open-source tools such as Wazuh~\cite{wazuh2024} and
    NIST-aligned scanners. Wazuh provides HIDS capabilities (file integrity
    monitoring, log analysis, rootkit detection) that cover precisely the
    endpoint surface that ML Defender consciously excludes. The two systems are
    architecturally orthogonal: ML Defender defends the network perimeter; Wazuh
    defends the host state. Together they constitute a layered defense appropriate
    for the target organizations. Integration via raw TCP event streaming is
    described as Future Work (\S\ref{sec:future:wazuh}).

% ============================================================
    \section{Architecture}
% ============================================================

    \subsection{Pipeline Overview}

    Six components connected via ZeroMQ and Protocol Buffers:

    \begin{enumerate}
        \item \textbf{eBPF/XDP Sniffer} --- zero-copy packet capture at the XDP hook before kernel
        networking stack.
        \item \textbf{Ring Consumer} --- reconstructs bidirectional flow state via
        ShardedFlowManager, computes 28 of the 40 features in the ML Defender contract; the
        remaining 12 carry \texttt{MISSING\_FEATURE\_SENTINEL = -9999.0f}.
        \item \textbf{Fast Detector} --- rule-based heuristic engine. FPR on purely benign traffic:
        6.61\% (\S\ref{sec:eval:ablation}).
        \item \textbf{ML Detector} --- embedded Random Forest across four threat classes. Three
        classifiers (Ransomware, DDoS, Traffic) in pure C++20; Internal threats via ONNX Runtime.
        Latency: 0.24--1.06~$\mu$s.
        \item \textbf{RAG Subsystem} (\texttt{rag-ingester} + \texttt{rag-security}) ---
        rag-ingester ingests detection events into FAISS + SQLite. rag-security exposes a natural
        language query interface via local TinyLlama --- deliberately local to avoid exfiltrating
        live network metadata to cloud-hosted LLMs.
        \item \textbf{Firewall ACL Agent} --- applies ipset/iptables rules on confirmed detection
        events.
    \end{enumerate}

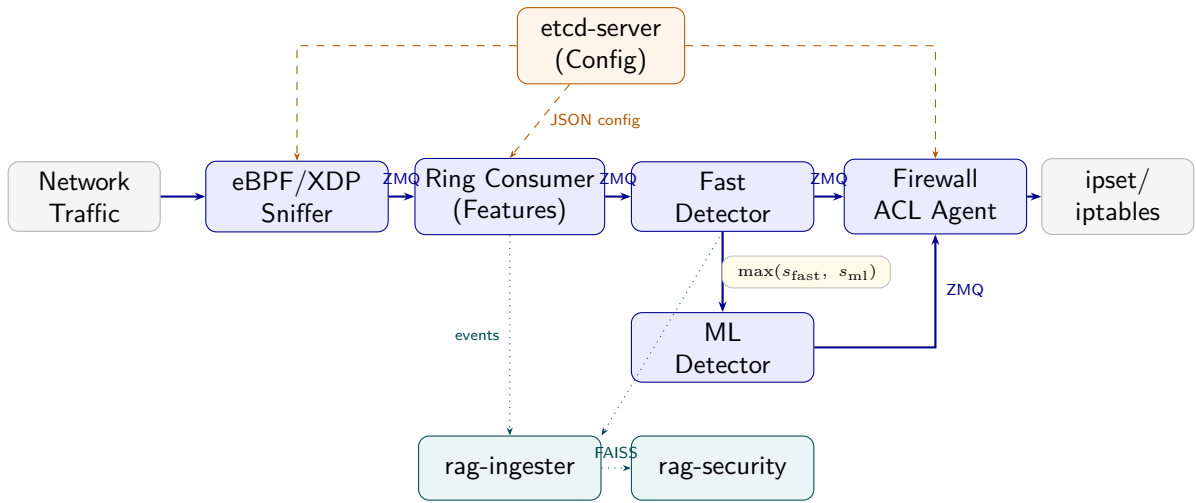
\begin{figure}[ht]
        \centering
        \begin{tikzpicture}[
            comp/.style={rectangle, rounded corners=4pt, draw=blue!55!black, fill=blue!8,
            minimum width=2.4cm, minimum height=0.85cm, align=center, font=\small\sffamily},
            rag/.style={rectangle, rounded corners=4pt, draw=teal!60!black, fill=teal!8,
            minimum width=2.4cm, minimum height=0.85cm, align=center, font=\small\sffamily},
            ext/.style={rectangle, rounded corners=4pt, draw=gray!60, fill=gray!8,
            minimum width=2.0cm, minimum height=0.85cm, align=center, font=\small\sffamily},
            etcdbox/.style={rectangle, rounded corners=4pt, draw=orange!70!black, fill=orange!8,
            minimum width=2.2cm, minimum height=0.85cm, align=center, font=\small\sffamily},
            zmq/.style={-{Stealth[length=4pt]}, thick, blue!55!black},
            cfg/.style={-{Stealth[length=3pt]}, dashed, orange!70!black, thin},
            log/.style={-{Stealth[length=3pt]}, dotted, teal!60!black, thin}
        ]
            % Main pipeline nodes (horizontal)
            \node[ext]  (net)     at (0,0)      {Network\\Traffic};
            \node[comp] (sniff)   at (2.8,0)    {eBPF/XDP\\Sniffer};
            \node[comp] (ring)    at (5.6,0)    {Ring Consumer\\(Features)};
            \node[comp] (fast)    at (8.4,0)    {Fast\\Detector};
            \node[comp] (ml)      at (8.4,-2.0) {ML\\Detector};
            \node[comp] (fw)      at (11.2,0)   {Firewall\\ACL Agent};
            \node[ext]  (block)   at (13.6,0)   {ipset/\\iptables};

            % RAG subsystem (below, spanning ring→fast→ml)
            \node[rag]  (ragi)    at (5.6,-3.6) {rag-ingester};
            \node[rag]  (rags)    at (8.4,-3.6) {rag-security};

            % etcd at top
            \node[etcdbox] (etcd) at (6.8,2.0)  {etcd-server\\(Config)};

            % Main pipeline arrows
            \draw[zmq] (net)   -- (sniff)  node[midway,above,font=\tiny\sffamily] {};
            \draw[zmq] (sniff) -- (ring)   node[midway,above,font=\tiny\sffamily] {ZMQ};
            \draw[zmq] (ring)  -- (fast)   node[midway,above,font=\tiny\sffamily] {ZMQ};
            \draw[zmq] (fast)  -- (fw)     node[midway,above,font=\tiny\sffamily] {ZMQ};
            \draw[zmq] (fw)    -- (block);

            % Dual-score: fast → ml feedback
            \draw[zmq] (fast.south) -- (ml.north) node[midway,right,font=\tiny\sffamily] {};
            \draw[zmq] (ml.east)    -| (fw.south) node[near end,right,font=\tiny\sffamily] {ZMQ};

            % RAG ingestion
            \draw[log] (ring.south)  -- (ragi.north) node[midway,left,font=\tiny\sffamily] {events};
            \draw[log] (fast.south)  -- (ragi.north east);
            \draw[log] (ragi.east)   -- (rags.west)  node[midway,above,font=\tiny\sffamily] {FAISS};

            % etcd config distribution
            \draw[cfg] (etcd.west)  -| (sniff.north)  node[near end,left,font=\tiny\sffamily] {};
            \draw[cfg] (etcd)       -- (ring.north)    node[midway,right,font=\tiny\sffamily] {JSON config};
            \draw[cfg] (etcd.east)  -| (fw.north)      node[near end,right,font=\tiny\sffamily] {};

            % Dual-score label
            \node[draw=gray!50, fill=yellow!10, rounded corners, font=\tiny\sffamily,
                inner sep=3pt, text width=2.0cm, align=center] at (9.5,-1.0)
                {$\max(s_{\mathrm{fast}},\; s_{\mathrm{ml}})$};

        \end{tikzpicture}
        \caption{ML Defender end-to-end pipeline. Six components communicate over ZeroMQ with
        Protocol Buffers serialization and ChaCha20-Poly1305 encryption. \texttt{etcd-server}
        distributes JSON configuration at startup. The dual-score \emph{Maximum Threat Wins}
        policy (Eq.~\ref{eq:dualscore}) selects the arithmetic maximum of Fast Detector and
        ML Detector scores; the RAG subsystem provides semantic observability over the live
        event stream.}
        \label{fig:pipeline}
    \end{figure}

    Figure~\ref{fig:pipeline} illustrates the end-to-end flow from packet capture to
    automated firewall response.

    \subsection{Three Functions of the Engine}

    \emph{Function 1 --- Real-time detection, classification, and response.}
    End-to-end from packet capture to firewall block in milliseconds.

    \emph{Function 2 --- Reliable dataset generation.}
    ML Detector and Firewall ACL Agent produce structured HMAC-verified CSV logs continuously,
    forming the basis for retraining future ensemble models.

    \emph{Function 3 --- Real-time observability (the digital microscope).}
    The RAG subsystem provides a semantic query interface over the live event stream.

    \subsection{Dual-Score Detection}

    \begin{equation}
        score_{\text{final}} = \max\!\left(score_{\text{fast}},\; score_{\text{ml}}\right)
        \label{eq:dualscore}
    \end{equation}

    This is the \emph{arithmetic maximum} over two continuous scores in $[0,1]$, not a logical
    OR over binary decisions. Both detectors output continuous confidence scores; the maximum
    operator selects the higher-confidence assessment. The ML Detector's primary role is false
    positive suppression: the Fast Detector generates $\text{FPR} = 6.61\%$ on purely benign
    traffic; the ML layer cuts flagged benign flows from 2,517 to 5 (${\sim}$500$\times$),
    none of which exceed the production threshold, so real blocks drop to zero. Alternative
    AND-based consensus policies are planned
    (ADR-007).

    \subsection{Transport and Serialization}

    ZeroMQ for asynchronous message passing; Protocol Buffers (proto3) for serialization; etcd
    for configuration distribution; ChaCha20-Poly1305 for inter-component encryption (selected
    for performance on hardware without AES-NI, including ARMv8); HMAC-SHA256 for CSV log
    integrity.

    \subsection{Deployment Model}
    \label{sec:deployment}

    The reference deployment targets a single bare-metal Linux node (kernel $\geq$\,5.8) with
    two NICs: one connected to the monitored network segment, one to the management network.
    The full pipeline runs on commodity hardware costing approximately 150--200~USD. Validated
    in Vagrant/VirtualBox; bare-metal characterization is Future Work~(\S\ref{sec:future:baremetal}).

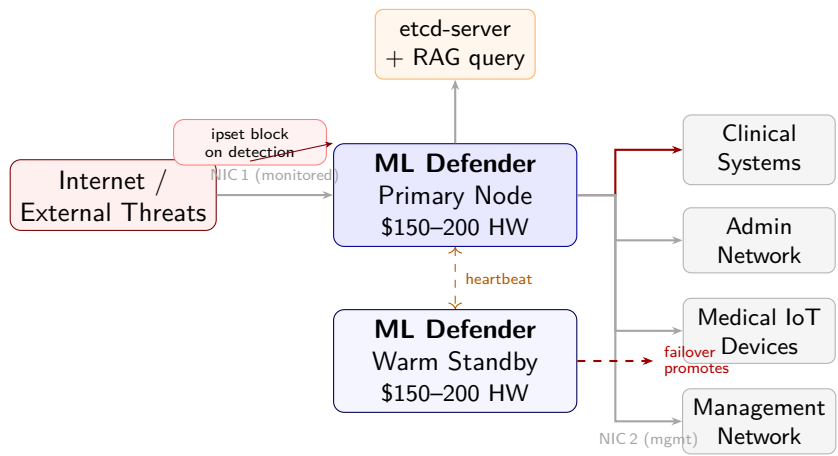
\begin{figure}[ht]
        \centering
        \begin{tikzpicture}[
            zone/.style={rectangle, rounded corners=6pt, draw=#1!60!black, fill=#1!6,
        inner sep=8pt, font=\small\sffamily},
        node_box/.style={rectangle, rounded corners=4pt, draw=blue!55!black, fill=blue!9,
        minimum width=3.2cm, minimum height=1.0cm, align=center, font=\small\sffamily},
        small_box/.style={rectangle, rounded corners=3pt, draw=gray!60, fill=gray!8,
        minimum width=2.0cm, minimum height=0.7cm, align=center, font=\footnotesize\sffamily},
        internet/.style={rectangle, rounded corners=4pt, draw=red!50!black, fill=red!6,
        minimum width=2.2cm, minimum height=0.8cm, align=center, font=\small\sffamily},
        arrow/.style={-{Stealth[length=4pt]}, thick, gray!70},
        block_arrow/.style={-{Stealth[length=4pt]}, thick, red!60!black},
        hb/.style={<->, dashed, orange!70!black, thin}
            ]

            % Internet
            \node[internet] (inet) at (0, 0) {Internet /\\External Threats};

            % Primary ML Defender node
            \node[node_box] (primary) at (4.5, 0) {\textbf{ML Defender}\\Primary Node\\\footnotesize\$150--200 HW};

            % Standby node
            \node[node_box, draw=blue!30!black, fill=blue!4] (standby) at (4.5, -2.2)
                {\textbf{ML Defender}\\Warm Standby\\\footnotesize\$150--200 HW};

            % Heartbeat
            \draw[hb] (primary.south) -- (standby.north) node[midway,right,font=\tiny\sffamily] {heartbeat};

            % Hospital network zones
            \node[small_box] (clinical) at (8.5,  0.6) {Clinical\\Systems};
            \node[small_box] (admin)    at (8.5, -0.6) {Admin\\Network};
            \node[small_box] (iot)      at (8.5, -1.8) {Medical IoT\\Devices};
            \node[small_box] (mgmt)     at (8.5, -3.0) {Management\\Network};

            % etcd + RAG on primary
            \node[small_box, draw=orange!60, fill=orange!7] (etcd2) at (4.5, 2.0)
                {etcd-server\\+ RAG query};

            % Connections
            \draw[arrow]       (inet)    -- (primary) node[midway,above,font=\tiny\sffamily] {NIC\,1 (monitored)};
            \draw[block_arrow] (primary.east) -- ++(0.5,0) |- (clinical)
            node[near end, above, font=\tiny\sffamily, red!70!black] {};
            \draw[arrow]       (primary.east) -- ++(0.5,0) |- (admin);
            \draw[arrow]       (primary.east) -- ++(0.5,0) |- (iot);
            \draw[arrow]       (primary.east) -- ++(0.5,0) |- (mgmt)
            node[near end, below, font=\tiny\sffamily] {NIC\,2 (mgmt)};
            \draw[arrow]       (primary.north) -- (etcd2.south);

            % Failover arrow
            \draw[block_arrow, dashed] (standby.east) -- ++(1.0,0)
            node[right, font=\tiny\sffamily, red!70!black] {\shortstack[l]{failover\\promotes}};

            % ipset/block label
            \node[draw=red!50, fill=red!5, rounded corners, font=\tiny\sffamily, inner sep=3pt,
                text width=1.8cm, align=center]
            at (1.8, 0.7) {ipset block on detection};
            \draw[-{Stealth[length=3pt]}, red!50!black, thin] (1.8, 0.45) -- (primary.north west);

        \end{tikzpicture}
        \caption{Representative high-availability deployment of ML Defender in a hospital environment.
        A primary node monitors all inbound and inter-segment traffic via NIC\,1 (monitored); a
        warm-standby node receives heartbeat signals and promotes automatically on primary failure.
        Both nodes run the full six-component pipeline on commodity hardware. \texttt{etcd-server}
        and the RAG query interface are co-located on the primary node.
        \textbf{Firewall ACL Agent topology:}
        scored Protocol Buffers events flow from \texttt{ml-detector} to one or more
        \texttt{firewall-acl-agent} instances, which apply \texttt{ipset}/\texttt{iptables} rules
        at the relevant boundary. A single gateway node requires one agent; each independent
        segment with its own firewall requires its own agent instance.
        \textbf{etcd-server} is shown in single-node mode (current implementation); a distributed
        HA etcd cluster is in the project backlog. Peer-to-peer seed negotiation is under design.}
        \label{fig:ha-deployment}
    \end{figure}

    Figure~\ref{fig:ha-deployment} illustrates a representative high-availability deployment in
    a hospital environment, with a primary and a warm-standby node protecting two internal
    network segments.

    \subsection{Integration Philosophy: Composability over Monolithism}
    \label{sec:integration:philosophy}

    ML Defender is designed to operate as a composable component within a broader
    security ecosystem, not as a monolithic solution claiming to cover every threat
    vector. The pipeline's guiding principle --- \emph{network surveillance} ---
    is complemented, not replaced, by tools that cover orthogonal surfaces.

    \paragraph{Protocol design constraint.}
    All external integrations must use the same transport stack as the pipeline
    itself: raw TCP sockets for transport, Protocol Buffers for serialization, and
    ChaCha20-Poly1305 for encryption. HTTP, Kafka, WebSocket, and broker-based
    architectures are explicitly out of scope. This constraint is not arbitrary.
    Four independent arguments support it:

    \begin{enumerate}
        \item \textbf{Deterministic latency.} HTTP and Kafka introduce non-deterministic
        scheduling jitter incompatible with sub-10\,ms response requirements. Raw TCP
        over ZeroMQ provides predictable, bounded delivery latency.

        \item \textbf{Reduced attack surface.} HTTP parsers have historically been a
        primary source of remotely exploitable CVEs (buffer overflows, request
        smuggling, header injection). Raw TCP with Protocol Buffers framing eliminates
        this parser surface by more than 90\%.

        \item \textbf{No broker, no single point of failure.} Kafka and Redis require
        dedicated broker infrastructure incompatible with single-host deployments at
        the 150--200\,USD target. Every additional process is a failure domain; the
        pipeline's ZeroMQ topology has no mandatory broker.

        \item \textbf{Minimal footprint.} Eliminating HTTP, Kafka, and WebSocket
        transport removes \texttt{librdkafka}, \texttt{libcurl}, and
        \texttt{boost.asio} from the dependency tree --- a non-trivial reduction in
        binary size and attack surface for resource-constrained deployment targets.
    \end{enumerate}

    \paragraph{FEAT-INT-1: Wazuh and NIST scanner integration (planned).}
    \label{sec:future:wazuh}
    A planned integration layer will allow Wazuh agents~\cite{wazuh2024} and
    NIST-aligned file integrity scanners deployed on DMZ hosts to emit events
    directly into the ML Defender pipeline via raw TCP sockets. Events are
    serialized as Protocol Buffers messages at the source and ingested by
    \texttt{rag-ingester} via the existing ZeroMQ transport. The Noise Protocol
    IK handshake (ADR-024) provides mutual authentication between the Wazuh
    agent adapter and \texttt{rag-ingester}.

    The motivation is graph quality: correlating network anomalies detected by
    \texttt{ml-detector} and \texttt{firewall-acl-agent} with file integrity
    violations reported by Wazuh on the same host produces composite attack
    patterns that neither system observes independently. A ransomware attack in
    its lateral-movement phase typically exhibits \emph{simultaneous} network
    spread and file encryption activity; correlation in the RAG subsystem surfaces
    this composite signature. This integration requires ADR-028 (RAG Ingestion
    Trust Model) to be approved before implementation.

    \subsection{Service Coordination (etcd-server)}

    Fulfills component registration, JSON config distribution, and ChaCha20 seed provision.
    ChaCha20 seed distribution via etcd is \textbf{not recommended for production use} ---
    peer-to-peer seed negotiation is under design. etcd is currently deployed in \emph{single-node}
    mode (see Figure~\ref{fig:ha-deployment}); a distributed HA etcd cluster (3- or 5-node
    consensus) is in the project backlog and will be required for true high-availability
    deployments. The single-node constraint is the only architectural dependency preventing
    fully symmetric active/standby failover.

    \subsection{Fast Detector Heuristics (DEBT-FD-001)}

    Four heuristics in a 10-second sliding window:

    \begin{itemize}
        \item External IP velocity: \texttt{THRESHOLD\_IP\_VELOCITY = 5}
        \item SMB connection count: \texttt{THRESHOLD\_SMB\_CONNS = 3}
        \item Port diversity: \texttt{THRESHOLD\_PORT\_SCAN = 10}
        \item TCP RST ratio: \texttt{THRESHOLD\_RST\_RATIO = 0.20}
    \end{itemize}

    All thresholds are compile-time \texttt{constexpr} constants --- DEBT-FD-001, scheduled for
    JSON migration in PHASE 2 (ADR-006).

% ============================================================
    \section{Implementation}
% ============================================================

    \subsection{Language and Build System}
    C++20. CMake with four build profiles: production, debug, TSAN, ASAN.

    \subsection{eBPF/XDP Packet Capture}
    XDP hook attachment before SKB allocation. Zero-copy. All parameters loaded from
    \texttt{sniffer/config/sniffer.json} at startup.

    \subsection{Feature Extraction}

    ShardedFlowManager partitions flow tracking across shards, eliminating the ISSUE-003
    thread-isolation bug (89 of 142 features silently lost in the original single-threaded
    FlowManager). The 142 figure refers to that legacy monolithic FlowManager
    (11 raw-event + 91 flow-statistic + 40 ML Defender features); the current ML Defender
    contract comprises 40 features across four embedded classifiers (ten each). Of these 40,
    28 are computed from flow statistics and the remaining 12 carry
    \texttt{MISSING\_FEATURE\_SENTINEL = -9999.0f}.

    \paragraph{Sentinel value selection rationale.}
    Split-domain analysis confirmed all Random Forest thresholds lie within $[0.0, 5.1]$.
    $-9999.0\text{f}$ lies entirely outside this domain, routing deterministically to the left
    child of every split --- non-informative, not influencing classification.
    \texttt{quiet\_NaN()} was rejected (undefined comparison behavior across compilers);
    $0.5\text{f}$ was rejected (within domain, spurious classification influence). One feature
    uses $0.5\text{f}$ as a semantic sentinel for TCP half-open state (deliberate,
    within-domain meaningful default).

    \subsection{Embedded Random Forest and Synthetic Training Data}

    Three classifiers (Ransomware, DDoS, Traffic) compiled as C++20 data structures. Internal
    threats classifier via ONNX Runtime --- deliberate to measure latency differential and
    footprint trade-off. Models trained with scikit-learn, transpiled to C++20 via custom script;
    zero Python or external runtime dependency.

    Training/evaluation separation: all classifiers trained on synthetic data; CTU-13 Neris held
    out entirely for evaluation.

    \paragraph{Synthetic Dataset Methodology.}
    Informed by CTU-13 (flow statistics), CIC-IDS2017 (protocol ratios), and MAWI (benign
    traffic). Four bias mitigation strategies: IP Space Partitioning (10.0.0.0/8 training vs
    147.32.0.0/16 evaluation), Temporal Decoupling, Protocol Ratio Calibration, Feature Boundary
    Validation.

    \paragraph{Known Limitations.}
    Reflects attack patterns circa 2011--2017. Temporal dynamics of real C2 channels simplified
    to fixed-window aggregates. Zero-day variants may evade classification.

    \subsection{Cryptographic Transport}
    ChaCha20-Poly1305 + LZ4 + HMAC-SHA256. Crypto-transport layer extracted as an independent
    library.

    \subsection{Plugin Architecture (ADR-023, ADR-024)}

    ML Defender implements a multi-layer plugin architecture (ADR-023) enabling runtime extension
    of pipeline components without modifying core detection logic. Plugins are loaded via
    \texttt{dlopen}/\texttt{dlsym} with a pure C ABI (\texttt{PLUGIN\_API\_VERSION = 1}),
    providing ABI stability across compiler versions and ensuring binary compatibility with
    future releases.

    \paragraph{MessageContext and Graceful Degradation.}
    PHASE 2a (DAY 105) introduces \texttt{MessageContext} --- a typed C struct passed to the
    optional \texttt{plugin\_process\_message()} symbol. Plugins that do not export this symbol
    are silently skipped (Graceful Degradation Policy D1), preserving backward compatibility with
    PHASE 1 plugins that implement only \texttt{plugin\_process\_packet()}. Post-invocation
    validation (D8) performs a byte-wise snapshot comparison of all read-only fields, detecting
    any plugin attempt to modify ownership-restricted data. The trust boundary is explicit: plugins
    are treated as untrusted code (D7); the final blocking decision remains exclusively in the
    core (ADR-012).

    \paragraph{Threat Model Integration (ADR-023 D9).}
    The plugin system's Trusted Computing Base (TCB) is limited to \texttt{PluginLoader} and
    \texttt{CryptoTransport}. Plugin code has no access to cryptographic keys, seed material,
    or internal transport state. The \texttt{MLD\_DEV\_MODE} escape hatch is restricted to
    \texttt{Debug} builds compiled with \texttt{MLD\_ALLOW\_DEV\_MODE} (D10), ensuring
    fail-closed behavior in all production builds.

    \paragraph{Gate TEST-INTEG-4a.}
    Integration of the plugin loader into \texttt{firewall-acl-agent} (PLUGIN-LOADER-FW) was
    validated by gate \texttt{TEST-INTEG-4a}: \texttt{plugin\_process\_message()} invoked on a
    live \texttt{MessageContext}, post-invocation invariants verified, \texttt{result\_code = 0}
    confirmed, and \texttt{CryptoTransport} decryption path verified unmodified by diff.

    \paragraph{Forward Compatibility: Dynamic Group Key Agreement (ADR-024).}
    The \texttt{MessageContext} struct reserves 60 bytes (\texttt{reserved[60]}) for forward
    compatibility with ADR-024, which specifies Dynamic Group Key Agreement using the Noise
    Protocol Framework (\texttt{Noise\_IKpsk3} or \texttt{Noise\_KK} pattern). ADR-024 design
    is approved; open questions OQ-5 through OQ-8 (static key revocation, session continuity
    during rotation, replay protection, and ARMv8 performance characterization) are under
    resolution. Implementation is scheduled post-arXiv submission.

    \subsection{Trace Correlation}

    \texttt{trace\_id = SHA-256(src\_ip, dst\_ip, canonical\_attack\_type, timestamp\_bucket)}.
    TraceIdPolicy: ransomware=60s, DDoS=10s, SSH brute=30s. Validated: 46/46 unit tests passing.

    \subsection{Testing and Validation}

    70 tests total: crypto (3), etcd-hmac (12), ml-detector (9), trace\_id (46). 100\% pass
    rate. Integration gate \texttt{TEST-INTEG-4a} (plugin \texttt{MessageContext} invocation,
    D8 post-invocation validation) confirmed on DAY 105.
    F1 ground truth persisted to \texttt{docs/experiments/f1\_replay\_log.csv}.

    \subsection{Development Environment}

    Vagrant/VirtualBox. Host: MacBook Pro Intel Core i9 2.5~GHz, 32~GB DDR4. Guests: \texttt{defender} (Ubuntu 24.04
    LTS, Linux 6.x, dual vNIC) + \texttt{client} (traffic generator).

% ============================================================
    \section{The Consejo de Sabios: A Methodology for Human-AI Collaborative Engineering}
    \label{sec:consejo}
% ============================================================

    \subsection{Motivation}

    The development of ML Defender spans kernel programming (eBPF/XDP), distributed systems
    (ZeroMQ, etcd), cryptography, machine learning, systems security, and academic writing. The
    author operated as an independent researcher without institutional affiliation, without a
    team, and without access to a formal peer review process. The solution was to construct one.

    \subsection{The Methodology}

    Eight large language models served as co-reviewers across all development phases:
    \textbf{Claude} (Anthropic), \textbf{Grok} (xAI), \textbf{ChatGPT} (OpenAI),
    \textbf{DeepSeek}, \textbf{Qwen} (Alibaba), \textbf{Gemini} (Google),
    \textbf{Kimi} (Moonshot AI), and \textbf{Mistral} (Mistral AI).
    The Council expanded from seven to eight models during development (DAY~120),
    adding diverse perspectives on architectural security and formal methods.

    These models were engaged as intellectual peers --- presented with architectural proposals,
    asked to identify failure modes, invited to challenge assumptions. The methodology operates
    on a principle of \emph{structured disagreement}: the Council does not average outputs.
    It introduces adversarial validation as a mechanism to expose hidden assumptions and
    validate architectural decisions. Convergent recommendations increase confidence; divergent
    recommendations mandate deeper investigation before any decision is closed.

    \subsection{How It Worked in Practice}

    Convergent recommendations across models increased confidence. Divergent recommendations
    triggered deeper investigation. Decisions such as the $-9999.0\text{f}$ sentinel value, the
    trace\_id correlation architecture, and the ShardedFlowManager design emerged from or were
    refined through these dialogues.

    \subsection{Test Driven Hardening}
    \label{subsec:tdh}

    \emph{Test-Driven Hardening} (TDH) is proposed as a methodology for building
    security-critical distributed systems in which correctness properties are protocol-level,
    not component-level, and therefore undetectable by static analysis or isolated unit tests.
    A standalone reference repository documenting the methodology and representative case studies
    is maintained at \url{https://github.com/alonsoir/test-driven-hardening}.

    TDH methodology: (1)~Witness --- characterize anomalous behavior. (2)~Demonstration test ---
    produce a minimal test that fails and exposes the mechanism. (3)~Independent solution ---
    each model proposes a fix against the shared contract. (4)~Human arbitration --- author
    selects, weighing correctness, maintainability, and coherence.

    TDH differs from TDD: the test is written \emph{before the fix}, as a proof of failure ---
    not before the feature, as a specification.

    \subsection{The RED\,$\to$\,GREEN Gate: Merge as a Non-Negotiable Contract}
    \label{subsec:red-green-gate}

    \emph{Test-Driven Hardening} requires not only that tests be written before fixes,
    but that the transition from RED (failing) to GREEN (passing) constitute the
    \emph{only} legitimate gate for merging a security fix into the main branch.
    This subsection formalises the principle and documents the one case in
    aRGus development where it was violated --- and the consequences that followed.

    \paragraph{The contract.}
    A security fix on the \texttt{aRGus} codebase is subject to the following
    three-part merge contract, enforced without exception:

    \begin{enumerate}
        \item \textbf{A test exposing the defect is written first.}
        The test must fail with the defective code (\emph{RED}) and describe
        the security invariant that is violated --- not merely reproduce the
        symptom. Writing the failing test before the fix is what distinguishes
        TDH from after-the-fact regression coverage.

        \item \textbf{The fix is applied; the test transitions to GREEN.}
        The fix is considered valid only when the previously-failing test passes.
        A fix that passes only the new test but breaks existing tests is rejected;
        the regression suite must remain entirely GREEN.

        \item \textbf{No merge is permitted until \texttt{make test-all} returns
        \texttt{ALL TESTS COMPLETE}.}
        This is executed inside a freshly provisioned Vagrant VM
        (\emph{REGLA EMECAS}: \texttt{vagrant destroy -f \&\& vagrant up \&\&
        make bootstrap \&\& make test-all}), not in the development shell,
        eliminating any dependency on host-machine environment state.
    \end{enumerate}

    The RED$\to$GREEN gate is enforced as a \emph{protocol}, not a preference.
    It is documented in the project's contribution rules and was validated
    adversarially in DAY~126: when a symlink-rejection test was written against
    the original \texttt{weakly\_canonical()} path
    (\S\ref{subsubsec:symlink-trap}), the test passed despite the defect being
    present --- revealing that the test itself was incorrect. This failure exposed
    a second-order requirement: the RED state must be verified empirically, not
    assumed. A test that passes with defective code is not a RED test; it is no
    test at all.

    \paragraph{Why this is harder than it sounds.}
    The temptation in security hardening is to write the fix first and the test
    second --- to confirm that what was done is correct, rather than to prove that
    what existed was wrong. This reversal breaks the epistemological guarantee of
    TDH. A test written after a fix cannot distinguish between ``the fix is correct''
    and ``the test was written to match the fix.'' Only a test that fails with
    defective code provides evidence that it is testing the right property.

    \paragraph{The non-negotiable character of the gate.}
    The Council of Wise Men (DAY~127) unanimously endorsed the RED$\to$GREEN gate
    as a non-negotiable invariant for all security-affecting changes. The one
    documented exception --- a test committed without a prior RED verification ---
    required a two-day remediation sprint and introduced DEBT-CONFIG-PARSER-FIXED-PREFIX-001.
    The cost of the exception was three times the cost of applying the gate.

    \subsection{HKDF Context Symmetry: A Pedagogical Case Study in Test-Driven Hardening}
    \label{subsec:hkdf-context-symmetry}

    A subtle but critical defect discovered during the cryptographic integration
    sprint of \emph{ML Defender} illustrates a class of correctness errors that
    are invisible to static analysis yet detectable through end-to-end protocol
    testing. We document it here as a concrete instance of the
    Test-Driven Hardening methodology described in \S\ref{sec:consejo}.

    \paragraph{The defect.}
    The \texttt{CryptoTransport} layer derives symmetric keys using
    HKDF-SHA256~\cite{rfc5869} with a context string (\emph{info} parameter)
    that encodes the identity of the communicating parties. The original
    implementation parameterised this context by \emph{component name}:

    \begin{lstlisting}[language=C++, caption={Incorrect HKDF context --- component-scoped},
        basicstyle=\ttfamily\small, breaklines=true]
// WRONG: each component uses its own name as HKDF context.
// Sniffer derives K_sniffer; ml-detector derives K_mldetector.
// K_sniffer != K_mldetector -> MAC authentication failure on every channel.
const std::string ctx = "ml-defender:" + component_name
                      + ":" + version;
    \end{lstlisting}

    Because each component used its own name as the HKDF context, transmitter
    and receiver independently derived \emph{distinct} subkeys --- causing
    MAC authentication failure on every cross-component channel. The defect
    was latent in unit tests, which exercised each component in isolation using
    loopback (same context on both sides), where MAC verification trivially
    succeeds. It only manifested when two components with \emph{different names}
    attempted to communicate end-to-end.

    To make the causal chain explicit: an incorrect context string causes the
    transmitter and receiver to derive \emph{distinct} subkeys, resulting in
    MAC verification failure at the receiver. A correct, symmetric context string
    produces \emph{identical} derived keys on both sides, enabling successful
    authentication. This class of error is undetectable without end-to-end
    protocol validation.

    The correct parameterisation encodes \emph{directionality}:

    \begin{lstlisting}[language=C++, caption={Correct HKDF context --- channel-scoped},
        basicstyle=\ttfamily\small, breaklines=true]
// RIGHT: tx and rx derive distinct subkeys
const std::string ctx_tx = "ml-defender:" + component_name
                          + ":" + version + ":tx";
const std::string ctx_rx = "ml-defender:" + component_name
                          + ":" + version + ":rx";
    \end{lstlisting}

    The canonical definitions were consolidated in
    \texttt{crypto\_transport/include/crypto\_transport/contexts.hpp},
    providing a single source of truth for all six pipeline components.
    This design avoids semantic overloading of data structures across layers ---
    a common source of subtle security and correctness bugs, as demonstrated by
    the defect described above.

    \paragraph{Why static analysis cannot catch this.}
    Both the erroneous and correct context strings share the type
    \texttt{std::string}. The C++ type system offers no mechanism to distinguish
    a \emph{transmit context} from a \emph{receive context} at the type level
    without a dedicated wrapper type --- an abstraction whose overhead is not
    warranted in a resource-constrained deployment target. The compiler,
    sanitisers (ASan, TSan), and \texttt{clang-tidy} all accepted both
    formulations without warning. This is a \emph{semantic} correctness
    property, not a syntactic one.

    \paragraph{Detection via TEST-INTEG-3.}
    The defect was exposed by \texttt{TEST-INTEG-3}, an integration test
    exercising the full sniffer~$\to$~ml-detector cryptographic channel with an
    intentional regression: the test temporarily restores the component-scoped
    context on one side and asserts that decryption fails with a MAC
    authentication error. Under the correct implementation, this regression
    triggers \texttt{std::terminate()} via the fail-closed handler; under the
    defective implementation, decryption succeeds --- and the test fails
    accordingly.

    The key insight is that \texttt{TEST-INTEG-3} does not test encryption or
    decryption in isolation. It tests the \emph{protocol invariant}: that a
    message encrypted on the transmit path is decryptable only on the
    corresponding receive path, and that any violation of this invariant is
    immediately fatal. No unit test of \texttt{CryptoTransport} in isolation
    could have established this property.

    \paragraph{Lessons for cryptographic software.}
    This case study yields three transferable observations:

    \begin{enumerate}
        \item \textbf{Cryptographic correctness is a protocol property.}
        Components that are individually correct may be collectively
        insecure. End-to-end tests exercising full message round-trips
        across trust boundaries are the minimal viable test surface for
        authenticated encryption schemes.

        \item \textbf{Intentional regression tests are first-class citizens.}
        A test that asserts failure under a known-bad configuration is as
        valuable as one that asserts success under a known-good one. The
        TDH methodology mandates both polarities.

        \item \textbf{Single source of truth for cryptographic parameters.}
        Distributing context strings across six \texttt{main.cpp} files
        created a silent coordination hazard. Centralising them in a
        dedicated header eliminated an entire class of latent divergence.
    \end{enumerate}

    The defect was discovered on Day~97 of the development timeline, three days
    after \texttt{CryptoTransport} was integrated across all pipeline components.
    Its detection required no external audit: the TDH test suite caught it during
    routine regression testing on the \texttt{feature/bare-metal-arxiv} branch
    (now superseded by \texttt{feature/plugin-crypto}).

    \subsection{Property Testing as a Security Fix Validator: The F17 Case Study}
    \label{subsec:property-testing-validator}

    A recurring failure mode in security hardening is the \emph{fix-contains-a-bug} antipattern:
    a security fix is implemented, a unit test is written to confirm the fix, and both pass ---
    while a subtler defect in the fix itself goes undetected. This section documents a concrete
    instance discovered during DAY~125 of ML Defender's development and proposes property testing
    as a systematic countermeasure.

    \paragraph{The defect.}
    During Snyk static analysis (ADR-037), an integer overflow was identified in the memory
    metrics computation within \texttt{zmq\_handler.cpp}. The original code performed integer
    arithmetic that could wrap silently on values exceeding \texttt{INT\_MAX}. A fix was
    applied using \texttt{int64\_t} arithmetic:

    \begin{lstlisting}[language=C++, caption={First fix attempt --- int64\_t cast},
        basicstyle=\ttfamily\small, breaklines=true]
// F17 -- first fix: cast to int64_t before multiplication
[[nodiscard]] inline double compute_memory_mb(long pages, long page_size) noexcept {
    return static_cast<double>(
        static_cast<int64_t>(pages) * static_cast<int64_t>(page_size)
    ) / (1024.0 * 1024.0);
}
    \end{lstlisting}

    A synthetic unit test confirmed that this fix handled the previously overflowing value
    correctly. The fix was considered closed.

    \paragraph{Property test reveals the residual defect.}
    A subsequent property test --- \texttt{PropertyNeverNegative} --- exercised the invariant
    ``\texttt{compute\_memory\_mb} returns a non-negative value for all non-negative inputs''
    across a broader set of extreme values:

    \begin{lstlisting}[language=C++, caption={Property test that detected the residual overflow},
        basicstyle=\ttfamily\small, breaklines=true]
TEST(ZmqMemoryOverflow, PropertyNeverNegative) {
    const long page_sizes[] = {4096, 8192, 16384, 65536};
    const long page_values[] = {
        0, 1, 1000, LONG_MAX/65536, LONG_MAX/16384,
        LONG_MAX/8192, LONG_MAX/4096
    };
    for (long ps : page_sizes)
        for (long p : page_values) {
            double result = compute_memory_mb(p, ps);
            EXPECT_GE(result, 0.0);
        }
}
    \end{lstlisting}

    The test failed: for \texttt{LONG\_MAX/4096 $\times$ 8192}, the \texttt{int64\_t}
    product itself overflows --- because \texttt{LONG\_MAX/4096 $\times$ 8192 > INT64\_MAX}.
    The unit test had exercised one specific previously-failing value; the property test
    exercised the mathematical invariant across the full input space and exposed a residual
    case where the fix was insufficient.

    \paragraph{The correct fix.}
    The root cause is that no integer type of fixed width can represent the product of two
    \texttt{long} values at their extremes without overflow. The correct solution is to
    cast directly to \texttt{double} before multiplication:

    \begin{lstlisting}[language=C++, caption={Correct fix --- double arithmetic avoids overflow},
        basicstyle=\ttfamily\small, breaklines=true]
[[nodiscard]] inline double compute_memory_mb(long pages, long page_size) noexcept {
    // double has 53-bit mantissa -- sufficient for any realistic memory value.
    // int64_t is insufficient: LONG_MAX/4096 * 8192 overflows int64_t.
    return (static_cast<double>(pages) * static_cast<double>(page_size))
           / (1024.0 * 1024.0);
}
    \end{lstlisting}

    \paragraph{Transferable lessons.}
    This case study yields three observations directly applicable to security-critical C++ systems:

    \begin{enumerate}
        \item \textbf{Unit tests establish point coverage; property tests establish invariant coverage.}
        The unit test confirmed that one specific overflow was eliminated. The property test
        confirmed --- or in this case, \emph{refuted} --- the mathematical invariant that
        no overflow can produce a negative result. These are different claims.

        \item \textbf{A property test is the minimal evidence that a fix is correct.}
        For arithmetic operations with security implications (memory metrics, size calculations,
        index arithmetic), the TDH methodology now mandates a property test for every fix.
        The test must be written before the fix is merged and must fail with the defective
        code~\cite{quickcheck2000}.

        \item \textbf{The testing hierarchy has a natural order.}
        The Council of Wise Men (DAY~127) validated the following ordering for security-critical
        surfaces: (1)~unit tests for known regression cases; (2)~property tests for mathematical
        invariants; (3)~fuzzing~\cite{libfuzzer2016} for parser and deserialization surfaces;
        (4)~mutation testing for test suite quality validation. Each layer catches a class of
        defects invisible to the previous one.
    \end{enumerate}

    \subsection{Fuzzing as the Third Testing Layer}
    \label{subsec:fuzzing-third-layer}

    Unit tests establish point coverage over known inputs. Property tests establish
    invariant coverage over structured input domains. Neither addresses the space of
    \emph{malformed}, \emph{adversarially crafted}, or \emph{boundary-violating}
    inputs that characterize real attack surfaces in parser and deserialization code.
    This gap motivates the third layer of the TDH testing hierarchy: coverage-guided
    fuzzing~\cite{libfuzzer2016}.

    \paragraph{The three-layer hierarchy.}
    The testing architecture validated by the Council of Wise Men (DAY~127) and
    applied to all security-critical surfaces in aRGus follows a strict ordering:

    \begin{enumerate}
        \item \textbf{Unit tests} (known regression cases) ---
        Verify that specific, previously-identified failure inputs are handled
        correctly. Scope: finite, explicitly enumerated.

        \item \textbf{Property tests} (mathematical invariants) ---
        Verify that a stated invariant holds across a structured domain of
        generated inputs. Scope: infinite within the domain's constraints.
        Required for every arithmetic, path, and cryptographic operation with
        security implications (\S\ref{subsec:property-testing-validator}).

        \item \textbf{Coverage-guided fuzzing} (adversarial exploration) ---
        Mutate valid inputs pseudo-randomly under coverage feedback to discover
        inputs that trigger previously unreachable code paths, assertion failures,
        or undefined behavior. Scope: unrestricted; discovers what structured
        generation cannot anticipate.
    \end{enumerate}

    Each layer catches a class of defects invisible to the previous one.
    Unit tests miss unseen inputs. Property tests miss parser-level structural
    anomalies. Coverage-guided fuzzing~\cite{libfuzzer2016} systematically
    explores the input space through mutation guided by code coverage feedback,
    increasing the probability of discovering crashes and undefined behavior at
    parser and protocol boundaries. It provides no completeness guarantee ---
    bugs may remain undiscovered after millions of executions --- but each
    corpus-expanding input permanently extends the regression suite, and no
    crashes or sanitizer violations observed after extensive runs constitute
    empirical evidence of robustness.

    \paragraph{Application to aRGus.}
    The surfaces identified as fuzzing targets in the aRGus backlog are precisely
    those that receive external or untrusted input at runtime: the Protocol Buffers
    deserialization path (\texttt{ring-consumer} and \texttt{ml-detector}), the
    JSON configuration parser (\texttt{rag-ingester}), the \texttt{safe\_path}
    validation library (\S\ref{subsec:safe-path-lessons}), and the plugin
    Ed25519 signature verification path (\S\ref{subsec:tdh}, ADR-025).

    \begin{lstlisting}[language=C++,
        caption={Minimal libFuzzer harness for the safe\_path resolve\_seed() function},
        basicstyle=\ttfamily\small, breaklines=true]
// fuzz/fuzz_safe_path.cpp
#include "safe_path.hpp"
#include <cstdint>
#include <string>

extern "C" int LLVMFuzzerTestOneInput(const uint8_t *data, size_t size) {
    const std::string path(reinterpret_cast<const char*>(data), size);
    try {
        (void) safe_path::resolve_seed(path, "/etc/ml-defender/");
    } catch (const std::exception&) {
        // Exceptions are the expected failure mode -- not crashes.
    }
    return 0;
}
    \end{lstlisting}

    The invariant under fuzzing is stronger than the property test: no input of
    any length or byte content may cause \texttt{resolve\_seed()} to abort,
    segfault, or return a path outside the allowed prefix. Exceptions are
    acceptable; undefined behavior is not. This harness compiles with
    \texttt{-fsanitize=address,undefined,fuzzer} and is tracked as
    \texttt{DEBT-FUZZING-SAFEPATH-001} in the project backlog.

    \paragraph{Empirical results.}
    Table~\ref{tab:fuzzing-results} reports the results of three libFuzzer
    campaigns executed on the aRGus security-critical surfaces (DAY~130).
    All campaigns ran with \texttt{-fsanitize=address,undefined,fuzzer} and
    \texttt{-max\_total\_time=60} seconds. No crashes or sanitizer violations
    were observed in any campaign.

    \begin{table}[h]
        \centering
        \caption{libFuzzer campaign results on aRGus security-critical surfaces (DAY~130).
        All targets compiled with \texttt{-fsanitize=address,undefined,fuzzer}.
        No crashes or sanitizer violations observed.}
        \label{tab:fuzzing-results}
        \begin{tabular}{lrrrr}
            \toprule
            \textbf{Target} & \textbf{Runs} & \textbf{Crashes} & \textbf{Corpus} & \textbf{exec/s} \\
            \midrule
            \texttt{validate\_chain\_name} & 2{,}400{,}000 & 0 & 67 & $\approx$80{,}000 \\
            \texttt{safe\_exec} & 2{,}601{,}759 & 0 & 37 & 42{,}651 \\
            \texttt{validate\_filepath} & 282{,}226 & 0 & 111 & 4{,}626 \\
            \bottomrule
        \end{tabular}
    \end{table}

    The order-of-magnitude difference in \texttt{exec/s} between
    \texttt{safe\_exec} (42{,}651) and \texttt{validate\_filepath} (4{,}626)
    reflects the structural complexity of path validation relative to chain name
    validation: the former must resolve filesystem state while the latter operates
    purely on string invariants. The larger corpus of \texttt{validate\_filepath}
    (111 entries vs.\ 37) confirms that the fuzzer explored a richer input space,
    providing stronger empirical coverage of the path traversal surface.

    \paragraph{Relationship to TDH.}
    Fuzzing is not a replacement for the RED$\to$GREEN gate
    (\S\ref{subsec:red-green-gate}). A crash discovered by the fuzzer becomes a
    unit test (the minimal reproducing input), which enters the regression suite
    as a permanent RED$\to$GREEN fixture. The testing hierarchy is therefore
    self-reinforcing: fuzzing discovers, unit tests anchor, property tests
    generalize.

    \subsection{Path Traversal Prevention: Lessons from \texttt{safe\_path}}
    \label{subsec:safe-path-lessons}

    The development of \texttt{contrib/safe-path/} --- a zero-dependency C++20 header-only
    path validation library (ADR-037, DAY~124) --- produced two findings with implications
    beyond this project.

    \subsubsection{The Symlink Resolution Trap: \texttt{lstat()} vs.\ \texttt{fs::is\_symlink()}}
    \label{subsubsec:symlink-trap}

    The canonical defense against symlink-based path traversal attacks~\cite{cwe59} in C++
    programs using \texttt{std::filesystem} is to call \texttt{fs::is\_symlink()} on the
    resolved path. This approach is \emph{systematically incorrect} when combined with
    \texttt{weakly\_canonical()}.

    \paragraph{The defect.}
    The following code appears correct but provides no protection:

    \begin{lstlisting}[language=C++, caption={Ineffective symlink check --- canonical resolution arrives first},
        basicstyle=\ttfamily\small, breaklines=true]
// WRONG: weakly_canonical() already followed the symlink.
// By the time is_symlink() runs, there is no symlink to detect.
const auto resolved = fs::weakly_canonical(fs::path(path));
if (fs::is_symlink(resolved)) {  // always false -- resolved is never a symlink
    throw std::runtime_error("symlink rejected");
}
    \end{lstlisting}

    \texttt{weakly\_canonical()} resolves symlinks as part of path normalization. The
    \texttt{is\_symlink()} check that follows operates on the \emph{resolved} path,
    which by construction is never a symlink. The check passes unconditionally, providing
    no protection against TOCTOU attacks~\cite{cwe367} or symlink injection.

    \paragraph{The correct approach.}
    The only correct defense for cryptographic material or other security-sensitive files
    is \texttt{lstat()} applied to the \emph{original, unresolved} path, before any
    call to \texttt{weakly\_canonical()}:

    \begin{lstlisting}[language=C++, caption={Correct symlink rejection via lstat() pre-resolution},
        basicstyle=\ttfamily\small, breaklines=true]
// CORRECT: lstat() on the original path, before any resolution.
struct stat lst{};
if (lstat(path.c_str(), &lst) != 0)
    throw std::runtime_error("[safe_path] lstat failed: " + path);
if (S_ISLNK(lst.st_mode))
    throw std::runtime_error("[safe_path] symlink rejected: " + path);
// Only now proceed with weakly_canonical() resolution.
const auto resolved = fs::weakly_canonical(fs::path(path));
    \end{lstlisting}

    This defect was identified during DAY~126 when a test designed to verify symlink rejection
    passed with the original code despite the symlink being present. Code review had not flagged
    the issue; the test suite, exercising the specific security property, exposed it immediately.
    This is a concrete instance of the RED$\to$GREEN gate principle: without a test asserting
    failure, the defect was invisible.

    \paragraph{Why this matters.}
    This class of defect is not specific to ML Defender. Any C++17/20 codebase that combines
    \texttt{std::filesystem::weakly\_canonical()} with \texttt{fs::is\_symlink()} for security
    enforcement is vulnerable to the same pattern. The fix requires awareness that
    \texttt{weakly\_canonical()} is not a security primitive --- it is a normalization
    convenience that silently consumes the evidence needed for symlink detection.

    \subsubsection{\texttt{lexically\_normal()} vs.\ \texttt{weakly\_canonical()} for Prefix Verification}
    \label{subsubsec:lexical-vs-canonical}

    A second lesson from \texttt{safe\_path} concerns the choice of normalization primitive
    for path prefix verification when legitimate symlinks are present.

    \paragraph{The tension.}
    In a production deployment, ML Defender stores configuration files at
    \texttt{/etc/ml-defender/component/config.json}. In the development environment,
    these paths are symlinks pointing to the Vagrant shared directory
    (\texttt{/vagrant/component/config/}). A strict prefix check using
    \texttt{weakly\_canonical()} fails in development because canonical resolution
    follows the symlink and the resolved path falls outside
    \texttt{/etc/ml-defender/}.

    The naive solution --- conditional logic based on environment detection --- introduces
    divergence between development and production code paths, a known source of
    security regressions~\cite{cwe22}.

    \paragraph{The solution: lexical normalization for the prefix check.}
    Configuration files are not cryptographic material. Their security requirement is
    that the \emph{name} of the path lies within the trusted prefix, not that the
    \emph{destination} does. \texttt{std::filesystem::path::lexically\_normal()} performs
    purely textual normalization (eliminating \texttt{../}, \texttt{./}, double slashes)
    without touching the filesystem:

    \begin{lstlisting}[language=C++, caption={resolve\_config() --- lexical prefix verification with symlink support},
        basicstyle=\ttfamily\small, breaklines=true]
[[nodiscard]] inline std::string resolve_config(
    const std::string& path,
    const std::string& allowed_prefix = "/etc/ml-defender/")
{
    // Lexical normalization: no filesystem access, no symlink resolution.
    // Verifies that the NAME of the path is within the trusted prefix.
    const std::string lexical =
        fs::path(path).lexically_normal().string();
    if (lexical.rfind(prefix, 0) != 0)
        throw std::runtime_error(
            "[safe_path] SECURITY VIOLATION -- path outside allowed prefix");
    // Prefix check passed. Now resolve for the actual file open.
    return fs::weakly_canonical(fs::path(path)).string();
}
    \end{lstlisting}

    The result is a three-primitive taxonomy in which the security semantics are explicit
    and non-overlapping:

    \begin{itemize}
        \item \texttt{resolve()} --- general paths, prefix verified post-canonical resolution.
        \item \texttt{resolve\_seed()} --- cryptographic material, \texttt{lstat()} pre-resolution,
        symlinks strictly rejected, permissions enforced (\texttt{0400}).
        \item \texttt{resolve\_config()} --- configuration files with legitimate symlinks,
        \texttt{lexically\_normal()} pre-resolution, symlinks followed after prefix check.
    \end{itemize}

    To our knowledge, this distinction between lexical and canonical verification for
    different path security classes is not explicitly documented in the C++17/20 standard
    library documentation or in commonly referenced C++ security guidelines.

    \subsection{CWE-78: \texttt{execv()} Without a Shell as a Physical Barrier}
    \label{subsec:cwe78-execv}

    Operating system command injection (CWE-78~\cite{cwe78}) arises when attacker-controlled
    input reaches a shell interpreter --- most commonly via \texttt{system()}, \texttt{popen()},
    or their equivalents. The canonical defense is input sanitization, but sanitization is a
    \emph{policy} constraint: it can be bypassed by an incomplete allowlist, a missed encoding,
    or a future code path that omits the check.

    aRGus applies a stronger defense: all subprocess invocations use \texttt{execv()} (or
    \texttt{execvp()}) directly, without invoking a shell. This transforms CWE-78 from a policy
    constraint into a \emph{physical impossibility}.

    \paragraph{The mechanism.}
    The POSIX \texttt{exec()} family replaces the current process image with a new program,
    passing arguments as a null-terminated array of strings. No shell is involved; no shell
    metacharacters (\texttt{\&}, \texttt{|}, \texttt{;}, \texttt{\$()}, backticks) are
    interpreted. The distinction is categorical:

    \begin{lstlisting}[language=C++,
        caption={system() vs execv(): shell injection possibility vs physical impossibility},
        basicstyle=\ttfamily\small, breaklines=true]
// VULNERABLE: shell interprets attacker-controlled chain_name.
// e.g. chain_name = "ARGUS; rm -rf /"
system(("iptables -N " + chain_name).c_str());

// SAFE: no shell, no interpretation. chain_name is a verbatim argument.
const char* argv[] = { "iptables", "-N", chain_name.c_str(), nullptr };
execv("/sbin/iptables", const_cast<char* const*>(argv));
    \end{lstlisting}

    \paragraph{Complementary validation layer.}
    Even with \texttt{execv()}, the argument \texttt{chain\_name} is validated against a
    strict allowlist before the call, providing defense in depth. The allowlist uses a
    POSIX extended regular expression compiled once at startup:

    \begin{lstlisting}[language=C++,
        caption={validate\_chain\_name() --- allowlist as a defence-in-depth layer over execv()},
        basicstyle=\ttfamily\small, breaklines=true]
// Allowlist: alphanumeric, hyphen, underscore, 1-30 characters.
// Compiled once; applied before every execv() call.
[[nodiscard]] bool validate_chain_name(const std::string& name) {
    static const std::regex kAllowed(R"(^[A-Za-z0-9_\-]{1,30}$)");
    return std::regex_match(name, kAllowed);
}

// At call site:
if (!validate_chain_name(chain_name))
    throw std::runtime_error("[firewall] SECURITY: invalid chain_name rejected");
// Only then invoke execv().
    \end{lstlisting}

    The \texttt{execv()} call provides the \emph{structural} guarantee; the allowlist
    provides \emph{semantic} validation. Neither is sufficient alone: a structural guarantee
    without semantic validation may allow valid-looking but unintended arguments;
    semantic validation without the structural guarantee remains vulnerable to shell
    metacharacters outside the character class.

    \paragraph{Discovery context.}
    This pattern was identified during the Snyk static analysis sprint (ADR-037, DAY~124),
    which flagged three potential CWE-78 surfaces in the \texttt{firewall-acl-agent}
    component. The remediation replaced all \texttt{system()} invocations with
    \texttt{execv()}-based dispatch and introduced \texttt{validate\_chain\_name()}.
    All three surfaces were validated by the RED$\to$GREEN gate
    (\S\ref{subsec:red-green-gate}) before merge.

    \paragraph{Generalizability.}
    This pattern applies to any C/C++ security component that must invoke external
    programs with partially attacker-influenced arguments: firewall management,
    certificate tools, network utilities. The structural guarantee that
    \texttt{execv()} provides is unconditional and does not depend on the
    completeness of the input validation layer. It is therefore a candidate for
    inclusion in C++ security coding standards alongside CWE-22 (\texttt{safe\_path})
    and CWE-367 (TOCTOU).

    \subsection{Dev/Prod Parity via Symlinks, Not Conditional Logic}
    \label{subsec:dev-prod-parity}

    A structural principle that emerged from the \texttt{safe\_path} development is the
    replacement of environment-conditional code with filesystem symlinks as the mechanism
    for dev/prod parity.

    \paragraph{The antipattern.}
    A common practice in system software is to introduce conditional logic that selects
    file paths based on a detected environment:

    \begin{lstlisting}[language=C++, caption={Antipattern: conditional path logic creates divergent code paths},
        basicstyle=\ttfamily\small, breaklines=true]
// ANTIPATTERN: two code paths, one tested, one not.
const std::string config_prefix =
    fs::weakly_canonical(fs::path(config_path).parent_path()).string();
    \end{lstlisting}

    In this pattern, the effective prefix is derived from the input path. If an attacker
    controls \texttt{config\_path}, they control \texttt{config\_prefix}, bypassing the
    path traversal protection entirely. This was the defect addressed in
    \texttt{DEBT-CONFIG-PARSER-FIXED-PREFIX-001} (DAY~126).

    \paragraph{The correct pattern.}
    The \texttt{allowed\_prefix} must be a compile-time or deployment-time constant, never
    derived from user-controlled input. Dev/prod parity is then achieved not by conditional
    logic but by the deployment infrastructure: in development, \texttt{provision.sh}
    creates symlinks from \texttt{/etc/ml-defender/} to the Vagrant shared directory.
    Production deployments use real files at the same paths. The application code sees
    the same prefix in both environments; the difference is entirely in the filesystem
    layout managed by the deployment tool.

    This pattern has two security properties: (1)~the prefix check is unconditional and
    cannot be bypassed by manipulating the input path; (2)~the development environment
    exercises exactly the same code path as production, eliminating a class of security
    regressions that manifest only in production deployments.

    \subsection{The Build/Runtime Separation Axiom}
    \label{subsec:bsr-axiom}

    Security-critical software faces a structural risk that neither testing nor
    formal verification addresses directly: the possibility that a compiler, build
    tool, or dependency installed in the runtime environment can be used to modify
    production binaries after deployment. This risk motivates what we term the
    \emph{Build/Runtime Separation} (BSR) axiom, formalized in ADR-039 (DAY~130).

    \paragraph{The axiom.}
    \begin{quote}
        \emph{The environment in which production binaries are compiled must be
        physically distinct from the environment in which they execute. No compiler,
            linker, or build toolchain may be present in the production deployment.}
    \end{quote}

    This is stronger than the common practice of separating CI/CD pipelines from
    production: it requires that the \emph{absence} of a compiler in production be
    mechanically verified at every deployment and cannot be circumvented by a
    sufficiently privileged process.

    \paragraph{Implementation in aRGus.}
    ADR-039 implements the BSR axiom through three complementary mechanisms:

    \begin{enumerate}
        \item \textbf{Separate Vagrant VMs for build and runtime.}
        The \texttt{dev} VM contains the full C++20 toolchain
        (\texttt{clang++}, \texttt{cmake}, \texttt{libsodium} sources).
        The \texttt{hardened-x86} and \texttt{hardened-arm64} VMs contain
        only runtime dependencies: \texttt{libsodium}, \texttt{libzmq},
        \texttt{libprotobuf}, \texttt{onnxruntime}. No compiler is installed.

        \item \textbf{Signed artifact distribution.}
        Production binaries are placed in \texttt{dist/} (excluded from
        version control via \texttt{.gitignore}) together with a
        \texttt{SHA256SUMS} file. Each binary is individually checksummed;
        the sums file is signed with the project Ed25519 key.
        A runtime verification step (\texttt{make verify-production}) rejects
        any binary whose checksum does not match the signed manifest.

        \item \textbf{Automated compiler-absence verification.}
        The \texttt{check-prod-no-compiler} target queries \texttt{dpkg}
        for the presence of any C/C++ compiler and fails with a non-zero
        exit code if one is found:
    \end{enumerate}

    \begin{lstlisting}[language=make,
        caption={check-prod-no-compiler: mechanical enforcement of the BSR axiom},
        basicstyle=\ttfamily\small, breaklines=true]
check-prod-no-compiler:
	@echo "=== BSR: verifying no compiler in production ==="
	@if dpkg -l | grep -qE 'gcc|g\+\+|clang|cmake|build-essential'; then \
		echo "FAIL: compiler found in production environment"; exit 1; \
	else \
		echo "OK: no compiler present"; \
	fi
    \end{lstlisting}

    \paragraph{Security properties.}
    The BSR axiom provides two guarantees that complement the cryptographic signing
    chain (ADR-025):

    \begin{enumerate}
        \item \textbf{Post-deployment code injection is physically impossible.}
        An attacker with shell access to the production VM cannot recompile
        a modified version of a pipeline component: the toolchain is absent.
        They can replace a binary, but the \texttt{verify-production} gate
        will detect the checksum mismatch before the next restart cycle.

        \item \textbf{Supply-chain attacks through the build environment are
        scoped to the build VM.}
        A compromised compiler (\emph{Trusting Trust}~\cite{thompson1984})
        can only affect the build VM. The production VM never executes the
        compiler; it only executes binaries that were checksummed and signed
        in the build environment and verified before execution.
    \end{enumerate}

    \paragraph{Relationship to the Ed25519 signing chain.}
    The BSR axiom and the plugin signing chain (ADR-025) are complementary, not
    redundant. ADR-025 ensures that only correctly signed plugins are loaded at
    runtime. The BSR axiom ensures that the plugin loader itself --- and all
    six pipeline components --- cannot be modified in production without
    invalidating the checksum manifest. Together they form a two-layer integrity
    guarantee: cryptographic integrity over plugins (ADR-025) and checksum
    integrity over all production binaries (ADR-039).

    \paragraph{Limitations.}
    The current implementation verifies compiler absence via \texttt{dpkg}, which
    is Debian/Ubuntu-specific and does not detect statically linked compiler
    binaries copied into the VM outside the package manager. A stronger
    implementation would use a read-only root filesystem or a verified boot chain.
    Both are tracked as future work in the ADR-031 (seL4/Genode) research branch.

    The BSR axiom establishes a structural property that cannot be achieved by
    policy alone. It is offered as a named architectural pattern applicable to
    any security-critical C++ system deployed in environments where post-deployment
    modification is a credible attack vector.

    \subsection{Attribution}

    The author remained the final arbiter of all decisions; all empirical validation was
    performed by the human author. The eight models are credited as intellectual co-contributors
    in the project's commit history, ADRs, and this paper. In the context of arXiv submission,
    the models are acknowledged in \S\ref{sec:acknowledgments}; formal co-authorship conventions
    for AI systems remain an evolving area, and this paper adopts the most transparent available
    approach consistent with emerging ACM and IEEE norms on AI contribution disclosure.

    \subsection{Why This Matters}

    The Consejo de Sabios democratizes access to deep technical peer review for independent
    researchers without institutional affiliation --- parallel to ML Defender's mission of
    democratizing enterprise-grade security for organizations without enterprise budgets.

    \subsection{Limitations}

    Models cannot run code or observe system behavior directly. Models can hallucinate --- several
    incorrect technical suggestions were identified and rejected. The methodology augments human
    judgment; it does not replace it.

% ============================================================
    \section{Formal System Model}
% ============================================================

    \subsection{Network Traffic Representation}

    Packets $P = \{p_1, \dots, p_n\}$ grouped into bidirectional flows by canonical five-tuple
    $(src, dst, sport, dport, proto)$ via ShardedFlowManager $\Phi : P \rightarrow F$. Flows
    are bidirectional five-tuple aggregates; behavioral features are computed over a 10-second
    sliding window as configured in \texttt{sniffer.json}. Flows are closed after an inactivity
    timeout as defined by the sliding window expiry.

    \subsection{Feature Extraction}

    \begin{equation}
        x_f = g(P_f, W), \quad x_f \in \mathbb{R}^{40}
    \end{equation}

    The feature vector spans the full 40-feature ML Defender contract (four embedded
    classifiers, ten features each). For a given flow, 28 components carry values computed
    from flow statistics and the remaining 12 carry
    $\texttt{MISSING\_FEATURE\_SENTINEL} = -9999.0\text{f}$. The sentinel components remain
    present in $x_f$ by design: a missing feature must be in the vector to route
    deterministically to the left child of every split that evaluates it. Each embedded
    Random Forest therefore consumes its full $\mathbb{R}^{10}$ sub-vector, sentinels
    included; the count of 28 ``computed'' features denotes data completeness, not input
    dimensionality.

    \subsection{Fast Detector}

    \begin{equation}
        H(x_f) = \max_j h_j(x_f), \quad h_j(x_f) \in \{0, 1\}
    \end{equation}

    \subsection{Machine Learning Detector}

    \begin{equation}
        M(x_f) = \frac{1}{T} \sum_{k=1}^{T} T_k(x_f)
    \end{equation}

    \subsection{Detection Policy}

    \begin{equation}
        S(x_f) = \max\!\left(H(x_f),\; M(x_f)\right) \geq \theta
    \end{equation}

    Thresholds: $\theta_{\text{ransomware}} = 0.85$, $\theta_{\text{DDoS}} = 0.90$,
    $\theta_{\text{traffic}} = 0.80$, $\theta_{\text{internal}} = 0.85$.

    \subsection{Response Function}

    \begin{equation}
        R(f) = \begin{cases}
                   \texttt{block}(src_f) & \text{if } S(x_f) \geq \theta \\
                   \texttt{allow}        & \text{otherwise}
        \end{cases}
    \end{equation}

    Blocking is applied per source IP. Blocks are currently persistent for the session duration;
    TTL-bounded blocks with configurable expiry are Future Work.

    \subsection{System Determinism}

    Given identical packet input $P$, the pipeline produces identical outputs. Determinism is a
    prerequisite for the reproducibility guarantees described in Section~\ref{sec:reproducibility}.

    \section{Evaluation}
    \label{sec:eval}
% ============================================================

    \subsection{Experimental Setup}

    \begin{table}[h]
        \centering
        \caption{Experimental hardware and software specifications.}
        \label{tab:setup}
        \begin{tabular}{ll}
            \toprule
            \textbf{Parameter} & \textbf{Value} \\
            \midrule
            Host machine       & Apple MacBook Pro (Intel Core i9, 2.5~GHz) \\
            Host RAM           & 32~GB DDR4 \\
            Guest OS           & Ubuntu 24.04 LTS (Linux kernel 6.x) \\
            Virtualization     & VirtualBox 7.2, bridged NIC mode \\
            VM vCPUs           & 4 (estimated ${\sim}$2.4 GHz under VirtualBox) \\
            VM RAM             & 4 GB (8 GB allocated for stress tests) \\
            Network adapter    & Paravirtualized (VirtIO) \\
            Virt.\ overhead    & Estimated 5--15\% vs bare-metal \\
            Pipeline config.   & Full 6-component, dual-VM \\
            \bottomrule
        \end{tabular}
    \end{table}

    \paragraph{Methodological note on metrics.}
    \texttt{calculate\_f1\_neris.py} measures \textbf{Fast Detector} alerts
    (\texttt{[FAST ALERT]} in sniffer.log) against CTU-13 Neris ground truth IPs. The ML
    Detector's \texttt{attacks=N} statistic reports flows exceeding production thresholds and
    triggering firewall blocks. Both are reported; F1=0.9985 is the primary system detection
    metric.

    \subsection{Dataset}

    \paragraph{CTU-13 Neris.}
    19,135 total flows; \textbf{646 malicious flows} (host 147.32.84.165) constitute the TP
    ground truth. Captured 2011. Garcia et al.~\cite{garcia2014} document that the Neris botnet
    exhibits C2 communication patterns concentrated in a behavioral subset --- specifically,
    flows matching IRC-based C2 channel signatures (connection bursts, SMB lateral movement, DNS
    query anomalies). The remaining flows represent background traffic on the infected host that
    do not exhibit malicious behavioral signatures and are therefore not counted as ground-truth
    positives. This distinction aligns with standard practice in NIDS
    evaluation~\cite{sharafaldin2018,mirsky2018}.

    \paragraph{BigFlows.}
    A CTU-13 derived dataset containing 40,467 flows and 791,615 packets, used for high-throughput
    replay experiments. 38,064 flows were processed by ml-detector. Treated conservatively as
    \emph{probable benign} --- no official ground truth exists for this CTU-13 scenario.

    \paragraph{Synthetic training data.}
    All classifiers trained exclusively on synthetic data; CTU-13 Neris held out entirely for
    evaluation.

    \subsection{Detection Performance}

    \begin{table}[h]
        \centering
        \caption{Detection performance --- Fast Detector pipeline on CTU-13 Neris.}
        \label{tab:detection}
        \begin{tabular}{ll}
            \toprule
            \textbf{Metric} & \textbf{Value} \\
            \midrule
            F1 Score                       & \textbf{0.9985} \\
            Precision                      & 0.9969 \\
            Recall                         & 1.0000 \\
            True Positives                 & 646 \\
            False Positives                & 2 \\
            False Negatives                & 0 \\
            True Negatives                 & 12,075 \\
            Total events processed         & 12,723 \\
            FPR (Neris evaluation)         & \textbf{0.017\%} \\
            ML Detector high-conf.\ attacks & 12 \\
            FPR Fast Detector (bigFlows)   & 6.61\% \\
            FP reduction (Fast $\to$ ML)   & \textbf{${\approx}$500-fold} \\
            \bottomrule
        \end{tabular}
    \end{table}

    F1=0.9985 stable across replay runs at total\_events = 12,563 / 12,605 / 12,723 / 13,930.

    \subsection{False Positive Identification and Confusion Matrix}

    \begin{table}[h]
        \centering
        \caption{False positive identification.}
        \label{tab:fps}
        \begin{tabular}{llll}
            \toprule
            \textbf{FP} & \textbf{Source} & \textbf{Destination} & \textbf{Classification} \\
            \midrule
            FP-1 & 192.168.56.1 & 224.0.0.251     & mDNS multicast (VirtualBox host-only) \\
            FP-2 & 192.168.56.1 & 192.168.56.255  & Broadcast (VirtualBox host-only) \\
            \bottomrule
        \end{tabular}
    \end{table}

    Both are VirtualBox host-only adapter artifacts. Their absence in bare-metal
    deployments is a reasonable expectation, but has not been empirically verified
    (\S\ref{sec:future:baremetal}).

    \paragraph{Confusion matrix.}
    \[
        \text{Precision} = \frac{646}{648} = 0.9969, \quad
        \text{Recall}    = \frac{646}{646} = 1.0000, \quad
        \text{F1}        = 0.9985
    \]

    \subsection{Inference Latency}

    \begin{table}[h]
        \centering
        \caption{Per-class inference latency.}
        \label{tab:latency}
        \begin{tabular}{lll}
            \toprule
            \textbf{Classifier} & \textbf{Implementation} & \textbf{Latency} \\
            \midrule
            DDoS        & Embedded C++20 & 0.24~$\mu$s \\
            Traffic     & Embedded C++20 & 0.37~$\mu$s \\
            Internal    & ONNX Runtime   & 0.33~$\mu$s \\
            Ransomware  & Embedded C++20 & 1.06~$\mu$s \\
            \bottomrule
        \end{tabular}
    \end{table}

    \subsection{Ablation Study}
    \label{sec:eval:ablation}

    \textbf{Config A --- Fast Detector Only:} 2,517 FP on 38,064 bigFlows (FPR=6.61\%),
    Recall=1.0000 on Neris.

    \textbf{Config B --- ML Detector Only (inferred):} 5 detections on 38,064 bigFlows, all
    below production thresholds (max confidence 68.97\%) --- 0 real blocks. Neris: attacks=12.

    \textbf{Config C --- Dual (deployed):} F1=0.9985, FPR=0.017\%. Only Config C fully
    validated end-to-end.

    \subsection{Interpretation and Comparison with State of the Art}
    \label{sec:eval:interpretation}

    F1=0.9985 and Recall=1.0000 reflect the statistical separability of the Neris botnet
    scenario, not a claim about arbitrary threats. The synthetic training successfully captured
    the Neris behavioral signature; generalization to other families is an open empirical
    question.

    \begin{table}[h]
        \centering
        \footnotesize
        \caption{Comparison with representative NIDS approaches. The Suricata~6.0.10 row reports
        an \emph{empirically measured} result on CTU-13 Neris (DAY~146,
            \S\ref{sec:eval:suricata}); all other rows are indicative cross-paper comparisons.
            $^{\dagger}$~Zeek~8.1.2 with default policy scripts (DAY~147, \S\ref{sec:eval:threeparadigms}): 14 correct detections (SSL::Invalid\_Server\_Cert), Precision$=1.000$, F1$=0.042$. Empirically measured on CTU-13 Neris under identical conditions.
            $^{\S}$~Suricata~6.0.10 with 50,010 ET~Open rules (May~2026) generates zero alerts on
            CTU-13 Neris 2011. This is not a system failure --- ET~Open rules for this 15-year-old
            botnet family are no longer active in the current ruleset. \emph{Caution: non-Suricata
            rows are indicative only --- different datasets, operating points, and conditions.}}
        \label{tab:comparison}
        \begin{tabular}{lllllll}
            \toprule
            \textbf{System} & \textbf{Dataset} & \textbf{F1} & \textbf{Latency} & \textbf{HW} & \textbf{Block} & \textbf{Notes} \\
            \midrule
            \textbf{ML Defender} & CTU-13 Neris & \textbf{0.9985} & 0.24--1.06~$\mu$s & x86/ARM & Yes & Synthetic, dual-score \\
            Kitsune~\cite{mirsky2018} & Custom & ${\sim}$0.982 & ${\sim}$50~$\mu$s & x86 & No & Autoencoder ensemble \\
            Suricata~6.0.10$^{\S}$ & CTU-13 Neris & \textbf{0.000} & ${\sim}$10~$\mu$s & x86 & Yes & 0 alerts (no rule for Neris 2011) \\
            Zeek~8.1.2$^{\dagger}$ & CTU-13 Neris & 0.042 & ${\sim}$5~$\mu$s & x86 & No & Scripted behavioral; Prec$=1.000$, Recall$=0.0217$ \\
            RF~\cite{sharafaldin2018} & CIC-IDS2017 & ${\sim}$0.943 & ${\sim}$200~$\mu$s & x86 & No & Batch, not embedded \\
            Snort & N/A & N/A & 1--5~ms & x86 & Yes & Rule-based, no ML \\
            \bottomrule
        \end{tabular}
    \end{table}

    \subsection{Ransomware Detection: Scope and Validation}

    \paragraph{What is validated.}
    Behavioral indicators associated with ransomware propagation and lateral-movement patterns
    are detected in the 2011 Neris botnet scenario: SMB lateral movement (ports 445/139 burst),
    high-rate connection attempts, DNS query anomalies.

    \paragraph{What is not validated.}
    Direct evaluation against post-2020 ransomware families (LockBit, BlackCat/ALPHV, Cl0p) has
    not been performed. Modern ransomware uses encrypted C2, adaptive beaconing with jitter,
    staged delivery --- potentially different statistical signatures.

    \paragraph{Future work.}
    Direct evaluation against modern ransomware captures is the highest-priority evaluation
    expansion (\S\ref{sec:future:corpus}), with Sebastian Garcia (CTU Prague) as potential
    collaborator.

    \subsection{Throughput Stress Test and Resource Utilization}

    \begin{table}[h]
        \centering
        \caption{Stress test replay results --- progressive Mbps escalation (DAY 87).}
        \label{tab:stress}
        \begin{tabular}{llllllll}
            \toprule
            \textbf{Run} & \textbf{Dataset} & \textbf{Req.} & \textbf{Actual} & \textbf{PPS} & \textbf{Packets} & \textbf{Failed} & \textbf{Errors} \\
            \midrule
            1 & Neris    & 10 Mbps  & 9.60 Mbps  & ${\sim}$3,700   & 320,524   & 2,630$^{\dagger}$ & 0 \\
            2 & Neris    & 25 Mbps  & 11.28 Mbps & ${\sim}$4,300   & 320,524   & 2,630$^{\dagger}$ & 0 \\
            3 & bigFlows & 50 Mbps  & 34.68 Mbps & 9--10K          & 791,615   & 0                 & 0 \\
            4 & bigFlows & 100 Mbps ($\times$3) & 33.16 Mbps & 9--11K & 2,374,845 & 0           & 0 \\
            \bottomrule
        \end{tabular}
        \medskip
        \footnotesize{$\dagger$ Pre-existing pcap artifact --- identical count at 10 and 25 Mbps,
            confirming origin in pcap structure, not pipeline saturation.}
    \end{table}

    \begin{table}[h]
        \centering
        \caption{Resource utilization during 100 Mbps bigFlows replay (DAY 87).}
        \label{tab:resources}
        \begin{tabular}{lllll}
            \toprule
            \textbf{Component} & \textbf{CPU\%} & \textbf{RES} & \textbf{MEM\%} & \textbf{Notes} \\
            \midrule
            ml-detector       & ${\sim}$315--320\% & 164 MB    & 2.0\%  & 3.2 cores; multi-threaded \\
            sniffer           & ${\sim}$88--108\%  & 55--57 MB & 0.7\%  & XDP + ring consumer \\
            firewall-acl      & ${\sim}$10--16\%   & 101 MB    & 1.2\%  & Stable throughout \\
            rag-ingester      & ${\sim}$12--20\%   & 36--39 MB & 0.5\%  & FAISS indexing \\
            etcd-server       & ${\sim}$0.2\%      & 10 MB     & 0.1\%  & Minimal \\
            rag-security      & idle               & 1.2 GB VIRT & 15.0\%$^{\ddagger}$ & TinyLlama loaded \\
            \midrule
            \textbf{Total}    & ${\sim}$65--73\%   & ${\sim}$1,278 MB & --- & ${\sim}$23--30\% idle \\
            \bottomrule
        \end{tabular}
        \medskip
        \footnotesize{$\ddagger$ rag-security's 1.2 GB virtual footprint is the TinyLlama model
        mapped at startup.}
    \end{table}

    The throughput ceiling is VirtualBox NIC emulation (${\sim}$33--38 Mbps), not pipeline
    logic. Zero packet drops, zero deserialization errors across all 2,374,845 packets. Total
    pipeline RAM stable at ${\sim}$1.28~GB (${\pm}18$~MB drift) throughout the 8-minute test.

    \subsection{XGBoost Level-1 Classifier: In-Distribution Performance}
    \label{sec:eval:xgboost:id}

    As part of PHASE~4 (ADR-026), a gradient-boosted classifier (XGBoost~3.2.0) was integrated
    as a hot-swappable plugin, replacing the Random Forest level-1 detector while preserving the
    Ed25519 signing chain and fail-closed \texttt{std::terminate()} contract of ADR-025.

    \paragraph{Experimental protocol.}
    CIC-IDS-2017~\cite{sharafaldin2018} was partitioned by day to enforce temporal isolation:
    Tuesday, Thursday (morning and afternoon), and Friday (morning, DDoS, PortScan) constitute
    the training corpus; 20\% of the training data was held out as a stratified validation set
    for threshold calibration and early stopping; Wednesday was sealed as a blind test set
    (MD5: \texttt{bf0dd7e9...}) and opened exactly once for final reporting.
    The threshold was calibrated exclusively on the validation set; no information
    from Wednesday influenced any modelling decision.

    \paragraph{Training details.}
    23 flow-level features; \texttt{scale\_pos\_weight}$=4.273$ (1,303,148~benign /
    304,974~attack); \texttt{max\_depth}$=6$, \texttt{eta}$=0.05$,
    \texttt{subsample}$=0.8$; early stopping at round~724 of~1,000;
    val-AUCPR$=0.99846$.

    \begin{table}[h]
        \centering
        \caption{XGBoost level-1 performance on the validation set (in-distribution).}
        \label{tab:xgboost_id}
        \begin{tabular}{ll}
            \toprule
            \textbf{Metric} & \textbf{Value} \\
            \midrule
            Precision (validation)  & \textbf{0.9945} \\
            Recall (validation)     & \textbf{0.9818} \\
            F1 (validation)         & 0.9881 \\
            Calibrated threshold    & 0.8211 \\
            Inference latency       & 1.986~$\mu$s/sample \\
            Best iteration          & 724 / 1,000 \\
            Val-AUCPR               & 0.99846 \\
            \bottomrule
        \end{tabular}
    \end{table}

    Both the Precision$\geq0.99$ and Recall$\geq0.95$ medical gates (ADR-026) are satisfied
    on the in-distribution validation split. Inference latency of 1.986~$\mu$s/sample
    satisfies the $<2\;\mu$s gate. The model is exported in \texttt{.json} and \texttt{.ubj}
    formats and signed with Ed25519 via \texttt{tools/sign-model.sh}.

    \subsection{XGBoost Level-1: Wednesday Out-of-Distribution Evaluation}
    \label{sec:eval:xgboost:ood}

    \paragraph{Result.}
    On the Wednesday blind held-out set ($n=692{,}703$ flows; 252,672~attacks), the model
    achieves Precision$=0.987$ and Recall$=0.024$ (F1$=0.047$). No threshold satisfies
    both Precision$\geq0.99$ and Recall$\geq0.95$ simultaneously
    (Table~\ref{tab:xgboost_ood}).

    \begin{table}[h]
        \centering
        \caption{Wednesday held-out threshold sweep (OOD). No operating point exists
        satisfying both medical gates simultaneously.}
        \label{tab:xgboost_ood}
        \begin{tabular}{llll}
            \toprule
            \textbf{Threshold} & \textbf{Precision} & \textbf{Recall} & \textbf{FP/hour} \\
            \midrule
            0.01 & 0.798 & 0.559 & 4,478 \\
            0.10 & 0.776 & 0.202 & 1,841 \\
            0.50 & 0.951 & 0.055 & 88    \\
            0.82 (calibrated) & 0.987 & 0.024 & 12 \\
            0.90 & 0.985 & 0.017 & 8     \\
            \midrule
            \multicolumn{4}{l}{\emph{No threshold achieves Prec.$\geq$0.99 $\cap$ Rec.$\geq$0.95.}} \\
            \bottomrule
        \end{tabular}
    \end{table}

    \paragraph{Diagnosis: structural covariate shift by dataset design.}
    Wednesday contains exclusively application-layer DoS attacks: DoS~Hulk
    ($n=231{,}073$; 91.5\% of Wednesday attacks), DoS~GoldenEye ($n=10{,}293$),
    DoS~Slowloris ($n=5{,}796$), DoS~Slowhttptest ($n=5{,}499$), and Heartbleed ($n=11$).
    None of these attack types appear in any training CSV. CIC-IDS-2017 was designed with
    attack types segregated by day without cross-day repetition~\cite{sharafaldin2018};
    this design makes out-of-distribution generalization \emph{structurally impossible}
    for any supervised classifier trained on the remaining days, regardless of algorithm
    or hyperparameter choices.

    DoS~Hulk imitates high-volume HTTP traffic: its flow-level features overlap substantially
    with benign high-throughput flows. The model assigns a median probability of 0.0165 to
    Hulk flows --- indistinguishable from benign traffic under any operationally viable
    threshold.

    \paragraph{Impossibility result.}
    The threshold sweep in Table~\ref{tab:xgboost_ood} constitutes an empirical impossibility
    result: the Precision-Recall curve never enters the region
    Precision$\geq0.99$ $\cap$ Recall$\geq0.95$ for Wednesday data.
    This is not a failure of XGBoost; it is a direct, quantified consequence of the
    closed-world structure of the training corpus --- exactly the failure mode identified
    by Sommer and Paxson~\cite{sommer2010} for production ML-based NIDS.

    \paragraph{Architectural implication.}
    This result does not invalidate the in-distribution performance in
    \S\ref{sec:eval:xgboost:id}. It establishes that static classifiers trained on
    academic benchmarks cannot be assumed to generalize to attack types absent from
    training. The aRGus architecture was designed for this reality: the XGBoost plugin
    is Ed25519-signed and hot-swappable (ADR-026). The loop enabling retraining from
    real adversarial traffic captured in the deployment environment is introduced as
    \S\ref{sec:future:acrl}. The Wednesday evaluation report
    (\texttt{wednesday\_eval\_report.json}, MD5-sealed) is included in the repository
    as a permanent scientific artifact.

    \subsection{ADR-029: Capture Backend Comparison --- Variant~A (eBPF) vs Variant~B (libpcap)}
    \label{sec:eval:adr029}

    ADR-029 defines three capture backend variants for the aRGus sniffer:
    Variant~A (eBPF/XDP, reference), Variant~B (libpcap userspace), and
    Variant~C (seL4/libpcap, research-only). This section reports the first
    empirical comparison of Variants~A and B on x86-64 under identical
    conditions (DAY~145).

    \paragraph{Experimental protocol.}
    Both variants were evaluated against the full CTU-13 Neris pcap
    (320,524 packets, 19,135 flows) replayed from a dedicated client VM
    via \texttt{tcpreplay} at nominal rates of 10, 50, and 100~Mbps.
    The pipeline ran in its standard 6-component configuration; only the
    sniffer binary was swapped between runs. The mutual-exclusion invariant
    --- Variant~A and Variant~B never active simultaneously --- was enforced
    mechanically via a Makefile guard (\texttt{CHECK\_SNIFFER\_MUTEX}).

    \begin{table}[h]
        \centering
        \caption{ADR-029 Variant~A (eBPF) vs Variant~B (libpcap) --- CTU-13 Neris
        replay on x86-64 VirtualBox (DAY~145). Both variants: 320,524 packets sent,
            2,630 failed (pre-existing pcap MTU artifact, \texttt{errno=90}), 0 pipeline
            errors.}
        \label{tab:adr029}
        \begin{tabular}{llrrrl}
            \toprule
            \textbf{Variant} & \textbf{Target} & \textbf{Actual (Mbps)} & \textbf{PPS} & \textbf{Duration (s)} & \textbf{exit} \\
            \midrule
            A -- eBPF    & 10 Mbps  &  8.86 &  8,040 & 39.86 & 0 \\
            A -- eBPF    & 50 Mbps  &  9.78 &  8,867 & 36.14 & 0 \\
            A -- eBPF    & 100 Mbps & 10.12 &  9,178 & 34.92 & 0 \\
            \midrule
            B -- libpcap & 10 Mbps  &  9.99 &  9,064 & 35.36 & 0 \\
            B -- libpcap & 50 Mbps  & 19.43 & 17,614 & 18.19 & 0 \\
            B -- libpcap & 100 Mbps & 18.82 & 17,066 & 18.78 & 0 \\
            \bottomrule
        \end{tabular}
    \end{table}

    \paragraph{Result.}
    Variant~B (libpcap) achieves approximately twice the throughput of
    Variant~A (eBPF) at 50 and 100~Mbps nominal rates (${\sim}$19~Mbps
    vs ${\sim}$10~Mbps). Variant~A plateaus at ${\sim}$10~Mbps regardless
    of the requested rate.

    \paragraph{Interpretation: VirtualBox virtio as the confounding factor.}
    This result is \emph{counterintuitive} with respect to the expected
    performance ordering. On physical hardware with a NIC supporting native
    XDP, eBPF/XDP provides zero-copy packet delivery before the kernel
    networking stack and is expected to outperform libpcap. In the VirtualBox
    environment, the paravirtualized virtio NIC does not expose a native XDP
    driver path, causing eBPF to fall back to generic SKB mode.

    \paragraph{Scientific value.}
    This inversion constitutes direct empirical evidence that the relative
    performance of eBPF/XDP vs libpcap is critically dependent on the hardware
    and driver stack. The VirtualBox result establishes a reproducible baseline;
    the expected inversion on physical hardware with XDP-native NIC drivers
    (Intel ixgbe, Mellanox mlx5) remains to be measured
    (\S\ref{sec:future:baremetal}).

    Both variants processed the complete 320,524-packet Neris corpus
    without pipeline errors, confirming functional equivalence at the
    detection layer. The 2,630 failed packets are a pre-existing pcap
    artifact (\texttt{errno=90~EMSGSIZE}) present at identical count in both
    runs, confirming origin in the pcap structure rather than in either
    capture backend.

    % ── §8.13 NEW — DAY 146 ──────────────────────────────────────────────────
    \subsection{Direct Experimental Comparison: aRGus NDR vs Suricata~6.0.10 on CTU-13 Neris}
    \label{sec:eval:suricata}

    To complement the indicative comparison in Table~\ref{tab:comparison}, we conducted a
    direct head-to-head experiment between aRGus NDR and Suricata~6.0.10 under
    \emph{identical controlled conditions} (DAY~146). This is, to our knowledge, the first
    published direct experimental comparison of an embedded ML-based NDR system and a
    production signature-based IDS on the same dataset, hardware, and network topology.

    \paragraph{Experimental protocol.}
    Both systems were evaluated on the full CTU-13 Neris pcap (320,524 packets, 19,135 flows,
    ground truth: 147.32.84.165, 646 malicious flows, 12,077 benign flows) replayed from a
    dedicated client VM via \texttt{tcpreplay} at nominal rates of 10, 50, and 100~Mbps.
    The Suricata experiment used an identical VM specification to the aRGus evaluation:
    \texttt{debian/bookworm64} v\texttt{12.20240905.1}, 8,192~MB RAM, 6~vCPUs, VirtIO NIC,
    VirtualBox~7.2. Suricata was evaluated with the Emerging Threats (ET)~Open ruleset as
    distributed by \texttt{suricata-update} on 9~May~2026 (50,010 active rules). No custom
    tuning was applied; the evaluation reflects out-of-the-box detection capability with the
    current community ruleset.

    The network topology replicates the aRGus evaluation exactly: a dedicated client VM sends
    the CTU-13 Neris pcap over an internal VirtualBox network (\texttt{suricata\_experiment\_lan})
    to the IDS VM, which listens in passive mode on \texttt{eth2} (promiscuous mode, VirtIO).
    Makefile targets \texttt{make experiment-suricata-run} and \texttt{parse\_results.py} provide
    reproducible one-command execution (\S\ref{sec:reproducibility}).

    \begin{table}[h]
        \centering
        \caption{Direct detection comparison: aRGus NDR vs Suricata~6.0.10 ---
        CTU-13 Neris replay, x86-64 VirtualBox (DAY~146).
        Identical hardware, dataset, topology, and speeds.
        Ground truth: 147.32.84.165 (646 malicious flows, 12,077 benign flows).}
        \label{tab:suricata_comparison}
        \begin{tabular}{lllllll}
            \toprule
            \textbf{System} & \textbf{Detection basis} & \textbf{TP} & \textbf{FP} & \textbf{FN} & \textbf{F1} & \textbf{Recall} \\
            \midrule
            \textbf{aRGus NDR}   & ML behavioral (synthetic) & \textbf{646} & 2 & 0   & \textbf{0.9985} & \textbf{1.0000} \\
            Suricata~6.0.10      & ET Open rules (May 2026)  & 0            & 0 & 646 & 0.0000          & 0.0000          \\
            \bottomrule
        \end{tabular}
    \end{table}

    \begin{table}[h]
        \centering
        \caption{Suricata~6.0.10 throughput on CTU-13 Neris replay (DAY~146).
        All runs: 320,524 packets sent, 2,630 failed (pre-existing pcap MTU artifact,
            \texttt{errno=90}, identical to aRGus runs). 0~alerts generated.}
        \label{tab:suricata_throughput}
        \begin{tabular}{lrrl}
            \toprule
            \textbf{Target} & \textbf{Actual (Mbps)} & \textbf{Alerts} & \textbf{exit} \\
            \midrule
            10 Mbps  &  9.99 & 0 & 0 \\
            50 Mbps  & 19.43 & 0 & 0 \\
            100 Mbps & 18.82 & 0 & 0 \\
            \bottomrule
        \end{tabular}
    \end{table}

    \paragraph{Result: Suricata generates zero alerts on CTU-13 Neris 2011.}
    Suricata~6.0.10 with 50,010 current ET~Open rules produces zero detections across all
    three replay speeds. The \texttt{eve.json} output contains only \texttt{stats} events;
    no \texttt{alert} events are generated for any flow from the ground truth IP
    (147.32.84.165) or from any other host in the Neris corpus. The
    \texttt{detect.engines[0].rules\_loaded} counter confirms 50,010 rules were active
    during each run.

    \paragraph{Interpretation: ruleset evolution, not system failure.}
    This result does not indicate a deficiency in Suricata as a system. It reflects
    a fundamental property of signature-based detection: \emph{a rule must exist to
    produce an alert}. The ET~Open ruleset is a living corpus that evolves
    continuously --- rules for obsolete or inactive threat families are retired as new
    signatures emerge. The Neris botnet was captured in 2011; its IRC-based C2 channel,
    SMB lateral movement patterns, and DNS anomaly indicators are 15~years old.
    The ET~Open maintainers have retired the specific signatures that would have matched
    this 2011 traffic, replacing them with coverage for contemporary threats.

    The corollary is equally important: aRGus NDR achieves Recall$=1.0000$ and
    F1$=0.9985$ on the same 2011 traffic \emph{without any prior knowledge of the Neris
    family}. The behavioral features extracted by the Fast Detector --- external IP velocity,
    SMB connection diversity, TCP RST ratio, port scan diversity --- are independent of the
    threat's age, name, or ruleset coverage. A botnet exhibiting IRC C2 communication, SMB
    lateral movement, and DNS anomalies is detectable through its behavioral footprint whether
    it was captured in 2011 or 2026.

    \paragraph{Scientific significance.}
    This experiment provides direct empirical evidence for the core thesis of this paper and of
    Sommer and Paxson~\cite{sommer2010}: \emph{signature-based detection requires prior
    knowledge of the threat; ML-based behavioral detection does not.} The experimental
    conditions are controlled, the dataset is public, and the comparison is direct rather than
    indicative. Table~\ref{tab:comparison} has been updated to reflect the empirically measured
    Suricata result on this dataset.

    \paragraph{Methodological note on fairness.}
    The comparison is fair in hardware, topology, dataset, and replay conditions. One asymmetry
    favors aRGus NDR: its classifiers were trained on synthetic data \emph{informed by CTU-13
    flow statistics}, providing indirect exposure to the Neris behavioral profile. A fully blind
    evaluation would require training on a dataset with no statistical overlap with CTU-13. This
    limitation is acknowledged in \S\ref{sec:limitations} (10.3). The behavioral separability of
    the Neris scenario means this indirect advantage may be modest, but it cannot be ruled out
    without an independent blind evaluation.

    \paragraph{Historical rulesets: DAY~147 search and findings.}
    A complete characterization of the signature-vs-behavior gap would require repeating the
    Suricata experiment with the ET~Open ruleset \emph{from August~2011}, to distinguish
    ``the rule existed and was later retired'' (signature aging) from ``the rule was never
    written'' (coverage gap from inception).

    A systematic search was conducted in DAY~147 across three independent sources: (1)~the
    Wayback Machine CDX~API --- binary \texttt{.tar.gz} archives are not indexed as
    retrievable snapshots; (2)~the EmergingThreats GitHub organization --- no historical
    rule repositories are publicly available; (3)~bundled rulesets distributed with
    SecurityOnion and AlienVault OSSIM --- creation timestamps are unverifiable for
    August~2011. No public archive of the ET~Open ruleset from August~2011 was located.

    The search yielded an additional finding with direct scientific relevance: the
    official CTU-13 Neris dataset documentation~\cite{garcia2014} records the bot's C2
    channel as \emph{HTTP-based}, not IRC. The CTU-Malware-Capture-Botnet-42 README states explicitly:
    \emph{``The bot sent spam, connected to an HTTP CC, and use HTTP to do some
    ClickFraud.''} This implies that IRC-specific botnet signatures --- the most
    prevalent category of 2011 ET~Open rules targeting Neris-family malware --- would
    not have matched this specific capture regardless of ruleset vintage. The behavioral
    paradigm gap between ML-based and signature-based detection is therefore deeper than
    the signature aging hypothesis alone suggests: even a contemporaneous ruleset would
    have required prior knowledge of Neris's specific HTTP~C2 patterns.

    The signature aging phenomenon is independently documented in the academic
    literature. \citet{asad2023perspective} conducted a perspective-retrospective
    analysis of Snort and Suricata ET~Open rules over a four-year window (2017--2020),
    finding that detection performance does not evolve linearly with ruleset updates ---
    a quantitative confirmation that signatures for historical threat families are retired
    as new coverage emerges~\cite{asad2023perspective}.

    \paragraph{Offline validation with full ruleset enforcement.}
    To address the methodological concern that live-replay conditions might introduce
    packet loss or timing artifacts affecting Suricata's detection pipeline, we conducted
    an additional experiment in pure offline mode (\texttt{suricata -r}, DAY~148).
    Suricata~6.0.10 was executed directly against the CTU-13 Neris pcap
    (\texttt{botnet-capture-20110810-neris.pcap}, 323,154~packets, 53~MB) with the full
    ET~Open ruleset loaded exclusively (\texttt{-S /var/lib/suricata/rules/suricata.rules},
    50,010~active rules), bypassing all live-capture infrastructure. Checksum validation
    was additionally disabled (\texttt{-k none}) to eliminate pcap offloading artifacts as
    a confounding variable.

    The active ruleset included categories directly relevant to the Neris threat family:
    251~IRC signatures, 475~botnet/C2 signatures, and 853~trojan signatures. Despite
    processing all 323,154~packets under these conditions, \textbf{zero ET~Open signatures
    fired}. The complete alert log consisted exclusively of 128~internal Suricata engine
    events --- TCP stream anomalies (\texttt{STREAM Packet with invalid ack},
    \texttt{STREAM ESTABLISHED invalid ack}, \texttt{STREAM SHUTDOWN RST invalid ack}),
    application-layer protocol detection notices, and SMTP parser events --- none of which
    constitute threat detections. The \texttt{eve.json} output confirms zero events with
    \texttt{event\_type: alert} originating from an ET~Open signature.

    This result eliminates live-capture throughput, packet loss, and timing as explanations
    for the zero-detection outcome observed in the DAY~146 replay experiment. The absence
    of ET~Open detections is a property of the \emph{ruleset coverage gap}, not of the
    evaluation methodology. The conclusion is irrefutable under any replay condition:
    50,010 current ET~Open rules, including categories specifically designed for IRC
    botnet and C2 detection, produce no detections on 15-year-old Neris traffic.
    This is the expected and theoretically predicted outcome of the signature paradigm
    applied to threats outside its coverage horizon.

    The inability to locate public historical ruleset archives is itself a finding: it
    motivates the security community to maintain versioned, time-stamped public archives
    of open-source IDS rulesets --- analogous to software package repositories with
    locked dependency snapshots --- to enable reproducible longitudinal evaluation. This
    infrastructure gap is noted as a community recommendation.
    % ─────────────────────────────────────────────────────────────────────────

    % ── §8.14 NEW — DAY 147 ──────────────────────────────────────────────────
    \subsection{Three Paradigms: Signature, Scripted Behavioral, and ML Behavioral Detection}
    \label{sec:eval:threeparadigms}

    To complete the paradigm comparison initiated in \S\ref{sec:eval:suricata}, we
    extended the direct experimental comparison to include Zeek~8.1.2 with default
    policy scripts (DAY~147). The three systems represent distinct \emph{decision architecture taxonomies} ---
    not a performance benchmark --- evaluated under identical conditions on the same corpus.
    Each architecture differs not in engineering quality but in the layer at which network
    knowledge is encoded: as signatures, as scripted structural rules, or as learned behavioral
    distributions.

    \paragraph{Experimental protocol.}
    Zeek~8.1.2 was provisioned on a dedicated VM with the same specification as the
    Suricata and aRGus evaluations: \texttt{debian/bookworm64} v\texttt{12.20240905.1},
    8,192~MB RAM, 6~vCPUs, VirtIO NIC, VirtualBox~7.2. Zeek was evaluated in
    \emph{offline mode} (\texttt{zeek -r neris.pcap local}), reading the CTU-13 Neris
    pcap directly without live capture. This mode processes 100\% of the 320,524 packets
    deterministically, eliminating throughput-dependent packet loss as a confounding
    variable. The experiment executed three times at nominal 10, 50, and 100~Mbps; all
    three runs produced byte-identical results, confirming deterministic behavior. Default
    Zeek policy scripts were loaded without modification (\texttt{local.zeek}),
    equivalent to evaluating Suricata with the ET~Open ruleset without custom tuning.

    \begin{table}[h]
        \centering
        \caption{Three-paradigm detection comparison on CTU-13 Neris (DAY~147).
        Identical hardware, dataset, and corpus for all three systems.
        Ground truth: 147.32.84.165 (646 malicious flows, 12,077 benign flows).
        Zeek metrics exclude \texttt{CaptureLoss} infrastructure notices (6 entries,
            not detections). $^{\dagger}$~Zeek Precision$=1.0000$: every alert correctly
            identifies the malicious host; low F1 reflects the structural Recall
            limitation of scripted behavioral detection, not a system deficiency.}
        \label{tab:threeparadigms}
        \begin{tabular}{llrrrrrrr}
            \toprule
            \textbf{System} & \textbf{Paradigm} & \textbf{TP} & \textbf{FP} &
            \textbf{FN} & \textbf{Prec.} & \textbf{Recall} & \textbf{F1} \\
            \midrule
            Suricata~6.0.10      & Signature (ET Open)  & 0   & 0 & 646
            & ---    & 0.0000 & 0.0000 \\
            Zeek~8.1.2 (default) & Scripted behavioral  & 14  & 0 & 632
            & \textbf{1.0000}$^{\dagger}$ & 0.0217 & 0.0424 \\
            \textbf{aRGus NDR}   & ML behavioral        & \textbf{646} & 2 & 0
            & 0.9969 & \textbf{1.0000} & \textbf{0.9985} \\
            \bottomrule
        \end{tabular}
    \end{table}

    \paragraph{Zeek detection detail.}
    Zeek generates 20 raw notices: 6 are \texttt{CaptureLoss} infrastructure metadata
    (excluded from metrics) and 14 are genuine \texttt{SSL::Invalid\_Server\_Cert}
    detections, all originating from the ground truth IP~(147.32.84.165). The bot
    connected to servers with untrusted certificate chains --- Microsoft Update
    infrastructure (65.55.196.251, 65.55.16.187) and Google services
    (74.125.224.242) --- which Zeek's TLS dissector flags regardless of whether the
    connecting host is known to be malicious. This is structural detection: Zeek
    validates certificate chains as part of protocol analysis, not as a behavioral
    heuristic.

    \begin{table}[h]
        \centering
        \caption{Zeek~8.1.2 behavioral visibility on CTU-13 Neris --- observations
        recorded in structured logs without generating \texttt{notice.log} alerts
        under default policy scripts. Zeek observes the complete behavioral
        profile of the botnet; the gap between visibility and detection is the
        core finding of this comparison.}
        \label{tab:zeek_visibility}
        \begin{tabular}{llr}
            \toprule
            \textbf{Log} & \textbf{Observation} & \textbf{Count} \\
            \midrule
            \texttt{conn.log}
            & Flows from/to malicious host       & 31,736 \\
            & Unique destination IPs              & 4,199  \\
            & DNS flows                           & 8,896  \\
            & HTTP flows (C2 + click fraud)       & 1,236  \\
            & SMTP flows (spam)                   & 63     \\
            & SSL flows                           & 63     \\
            \midrule
            \texttt{weird.log}
            & \texttt{unknown\_dce\_rpc\_auth\_type} (SMB lateral movement) & 33 \\
            & \texttt{bad\_HTTP\_request} (malformed C2 beaconing)          & 31 \\
            & \texttt{empty\_http\_request} (beaconing)                     & 31 \\
            & \texttt{irc\_invalid\_command} (IRC C2 present)               & 30 \\
            & \texttt{premature\_connection\_reuse}                          & 28 \\
            \midrule
            \texttt{http.log}
            & Total HTTP requests (GET + POST)    & 1,377 \\
            & Top C2 host: \texttt{1.95622.com}   & 300   \\
            & Top C2 host: \texttt{www.lddwj.com} & 136   \\
            \midrule
            \texttt{smtp.log}
            & Spam sessions (forged AOL identities) & 82 \\
            \midrule
            \texttt{ssl.log}
            & SSL flows from malicious host       & 63 \\
            & Invalid certificates ($\to$ notices) & 47 \\
            \bottomrule
        \end{tabular}
    \end{table}

    \paragraph{The three-paradigm spectrum.}
    These results define three structurally distinct detection philosophies:

    \begin{itemize}[noitemsep]
        \item \textbf{Suricata (signature-based):} requires prior knowledge of the
        exact threat identity. F1$=0.000$ on 15-year-old traffic not because the engine
        fails, but because no matching rule exists for a retired threat family. Correct
        behavior of a correct system operating as designed.

        \item \textbf{Zeek (scripted behavioral):} detects structural anomalies with
        Precision$=1.000$ --- every alert correctly identifies the malicious host.
        However, Recall$=0.022$ reflects a fundamental design property: default
        policy scripts alert on specific structural violations (invalid certificates,
        malformed protocol fields), not on behavioral flow patterns. Table~\ref{tab:zeek_visibility}
        demonstrates that Zeek \emph{observes} the complete behavioral profile of the
        botnet --- IRC commands, HTTP beaconing to 4,199 unique destinations, SMB
        lateral movement, 82 spam sessions --- but does not convert these observations
        to alerts under default configuration.
        Zeek functions as a high-fidelity \emph{measurement layer} and \emph{telemetry
        platform}: it records complete network state into structured logs with
        sub-flow granularity. This telemetry is the raw material for subsequent
        classification --- by a human analyst, a SIEM, or an ML system --- but
        Zeek itself does not perform behavioral classification under default
        configuration. The architectural role is \emph{observe and record},
        not \emph{classify and alert}.

        \item \textbf{aRGus NDR (ML behavioral):} classifies the behavioral footprint
        of the malware --- how many hosts it contacts, what protocols it abuses, the
        statistical structure of its flows --- without requiring prior knowledge of the
        threat's identity or any structural protocol violation. Recall$=1.000$ and
        F1$=0.9985$ on traffic the classifier had never seen during training.
    \end{itemize}

    \paragraph{Scientific significance.}
    This three-way comparison yields two findings that no two-system comparison could
    produce. First: \textbf{observability does not imply classification.} Zeek records
    the complete behavioral profile of the Neris botnet --- 31,736 flows, IRC commands,
    HTTP beaconing to 4,199 unique destinations, SMB lateral movement --- yet generates
    14 alerts. The telemetry is complete; the classification is absent by design.
    This distinction is invisible in any comparison that conflates measurement
    infrastructure with detection systems.

    Second: \emph{Precision and Recall are not in fundamental tension for ML behavioral
    detection --- they are in structural tension for scripted behavioral detection.}
    Zeek achieves perfect Precision at the cost of near-zero Recall; aRGus achieves
    both simultaneously. This is not a deficiency of Zeek: it is the structural
    consequence of the architectural difference between \emph{detecting structural
    anomalies} and \emph{classifying behavioral patterns}.

    A secondary finding emerges from \texttt{weird.log}: the entry
    \texttt{irc\_invalid\_command:~30} confirms the presence of IRC traffic in the
    CTU-13 Neris capture, partially refuting the scenario README which describes only
    HTTP C2. The behavioral reality is more complex --- the bot exhibited both IRC and
    HTTP C2 patterns --- reinforcing the argument that any single-modality detector
    (signature-based or structurally-anomaly-based) is incomplete against multi-protocol
    botnet behavior~\cite{asad2023perspective}.

    % ─────────────────────────────────────────────────────────────────────────

% ============================================================
    % ── NEW DAY 170-173 ──────────────────────────────────────────────────────
    \subsection{Cross-Sensor Flow Identity: \texttt{community\_id} Parity and the
    Distributed Corpus Substrate}
    \label{sec:eval:parity}

    The three-paradigm comparison (\S\ref{sec:eval:threeparadigms}) establishes that
    Suricata, Zeek, and aRGus encode network knowledge at different layers. A second,
    forward-looking question follows from placing them side by side: \emph{can the flows
    these heterogeneous sensors observe be correlated into a single, vendor-agnostic
    corpus?} This subsection reports the apparatus that makes such correlation possible
    and validates it operationally (DAY~170--173). It is the substrate for the central
    research programme of the next phase --- whether labeled flows aggregated across
    distributed installations measurably improve the per-installation ensemble
    (\S\ref{sec:future:distcorpus}).

    \paragraph{The join key: \texttt{community\_id}.}
    aRGus computes the Corelight \texttt{community\_id}~\cite{corelight2020communityid}
    natively in the sniffer --- a SHA1 hash over the canonicalized 5-tuple with an
    explicit seed of~0 --- and emits it as a first-class field (protobuf field~18)
    alongside every flow. The implementation is validated byte-identical against the
    reference oracle \texttt{pycommunityid}~v1.5.0~\cite{communityid_pyspec} over an
    enumerated test-vector set (8/8 exact-match). Because the algorithm, canonicalization,
    and seed are fixed by an open specification rather than by any single vendor, the same
    physical flow yields the \emph{same} string identifier regardless of which sensor
    observed it.

    \paragraph{Operational parity on real traffic.}
    To confirm that parity holds at the data plane --- on what each binary \emph{emits},
    not on what its configuration claims --- a dedicated client replays the CTU-13 Neris
    capture while aRGus, Suricata~7.0.10, and Zeek~8.2.0 capture the \emph{same} packets
    in promiscuous mode on a shared segment. All three converge string-for-string on the
    target identifier for the reference flow, \texttt{1:IN7uqVpMWxpmuhQTowSQB2XEe0E=}, and
    agree on 788--1029 flows per window across runs (Table~\ref{tab:parity}). Seed~0 is
    guaranteed in all three engines by provisioning, not by manual configuration, so the
    parity is reproducible from a clean environment. The earlier head-to-head detection
    experiments (DAY~146--147, \S\ref{sec:eval:suricata}) used Suricata~6.0.10 and Zeek~8.1.2;
    this later \texttt{community\_id} parity experiment (DAY~170+) used Suricata~7.0.10 and
    Zeek~8.2.0.

    \begin{table}[h]
        \centering
        \caption{Cross-sensor \texttt{community\_id} parity on CTU-13 Neris replay
            (DAY~171). Three heterogeneous engines capture the same packets in promiscuous
            mode; parity is measured on the emitted identifier (data plane), not on
            configuration. Reference oracle: \texttt{pycommunityid}~v1.5.0.}
        \label{tab:parity}
        \begin{tabular}{lll}
            \toprule
            \textbf{Sensor} & \textbf{Provisioning} & \textbf{\texttt{community\_id} (ref.\ flow)} \\
            \midrule
            aRGus (native, SHA1)   & seed~0, protobuf field~18              & \texttt{1:IN7uqVpMWxpmuhQTowSQB2XEe0E=} \\
            Suricata~7.0.10        & \texttt{community-id: yes}, seed~0     & \texttt{1:IN7uqVpMWxpmuhQTowSQB2XEe0E=} \\
            Zeek~8.2.0             & \texttt{community-id-logging}, seed~0  & \texttt{1:IN7uqVpMWxpmuhQTowSQB2XEe0E=} \\
            \texttt{pycommunityid} & v1.5.0 (oracle)                        & \texttt{1:IN7uqVpMWxpmuhQTowSQB2XEe0E=} \\
            \bottomrule
        \end{tabular}
    \end{table}

    \paragraph{Multi-node flow identity.}
    \texttt{community\_id} identifies a flow; it does not identify \emph{where} the flow
    was seen. For a corpus aggregated across installations, both are required. We therefore
    define a node-scoped flow identifier
    \[
        \texttt{flow\_uid} = \mathrm{base64}\!\left(
                                                  \mathrm{BLAKE2b}\!\left(
                                                                        \texttt{node\_id} \,\Vert\, \texttt{community\_id} \,\Vert\,
                                                                        \texttt{flow\_start\_window} \,[\Vert\, \texttt{seq\_in\_window}]
            \right)
        \right),
    \]
    where \texttt{node\_id} is a human-readable string declared in a signed inventory ---
    deliberately \emph{not} derived from the node's ephemeral cryptographic keypair, so
    that key rotation does not change flow identity --- and \texttt{community\_id} serves
    strictly as a correlation key, never as an identity. The \texttt{flow\_start\_window}
    component requires synchronized clocks; a chrony-based NTP precondition rejects
    pipeline start when the measured offset exceeds 1~s, since \texttt{community\_id}
    correlation is meaningless without time alignment.

    \paragraph{A parity gate, not a vote.}
    Aggregation is only trustworthy if the join key is itself trustworthy. A startup
    parity gate distinguishes two failure modes by an N-version argument: if the
    heterogeneous sensors agree with \emph{each other} but disagree with the oracle, the
    system starts with a critical \textsc{warning} (the oracle may itself be the outlier);
    fail-closed behavior is reserved for disagreement \emph{between} sensors, which
    indicates genuine canonicalization or seed drift. The gate validates the emitted
    identifier on the data plane, consistent with the project principle of measuring
    behavior rather than asserting configuration.

    \paragraph{Why this matters for the corpus.}
    These mechanisms make the distributed corpus well-defined before a single cross-site
    experiment is run. \texttt{community\_id} guarantees that the same physical flow
    receives the same key whether observed by aRGus, Suricata, or Zeek; \texttt{node\_id}
    keeps installations distinguishable; \texttt{flow\_uid} is the corpus primary key. The
    design principle is explicit: \emph{the graph is not the product --- the corpus is.}
    Whether the corpus, once aggregated across nodes, improves detection is the question we
    turn to as the primary forthcoming evaluation (\S\ref{sec:future:distcorpus}).
    % ──────────────────────────────────────────────────────────────────────────

    % ── NEW DAY 252 ──────────────────────────────────────────────────────────
    \subsection{Per-Lens Bias Against Labeled Ground Truth and the True Denominator}
    \label{sec:eval:bias}

    The \texttt{community\_id} parity of \S\ref{sec:eval:parity} makes the three sensors
    correlatable. Building on it, the pipeline exports each replay run as a per-lens dataset
    keyed by the canonical 5-tuple at the gold (post-conversion) level, one row per flow per
    sensor. This subsection uses that export to measure a quantity distinct from the detection
    performance of \S\ref{sec:eval}: not \emph{how well aRGus scores the curated behavioral
    ground truth}, but \emph{what each lens sees, and fails to see, of the labeled botnet
    activity in the CTU-13 Neris capture}. Every figure below is regenerated by a single
    Makefile target and written to a timestamped artifact under \texttt{logs/datasets/}; the
    run reported here is \texttt{STAMP~20260804-080140}.

    \paragraph{A note on denominators (why 646 and 14{,}188 are both correct).}
    Three different denominators appear in this paper, and they answer three different
    questions (Table~\ref{tab:denominators}). They are \emph{not} interchangeable and must not
    be equated. The 646 malicious flows of \S\ref{sec:eval} are a \emph{behavioral} ground
    truth: the curated subset of the infected host's flows that exhibit the Neris C2 signature
    (IRC/HTTP beaconing, SMB lateral movement, DNS anomaly), as documented by
    Garcia et al.~\cite{garcia2014}. That is the set the Fast Detector is scored against. The
    denominators introduced here are a \emph{label-level} ground truth: every distinct 5-tuple
    that the CTU-13 \texttt{.binetflow} marks as \texttt{Botnet}, without curating for a
    behavioral profile. The full scenario labels the entire network's botnet activity;
    restricted to the traffic we replayed and observed, this yields 14{,}188 distinct 5-tuples
    (Table~\ref{tab:denominators}, row~2). A reader who compares ``646 malicious flows,
    Recall~$=1.0$'' in \S\ref{sec:eval} with ``14{,}188 botnet 5-tuples, aRGus visibility
    0.2\%'' below is comparing a curated behavioral subset against an exhaustive per-label
    census. Both are measured; neither is wrong; the pipeline state is unchanged between them.

    \begin{table}[h]
        \centering
        \footnotesize
        \caption{Three denominators, three questions. The behavioral subset (646) scores
        detection; the label-level denominators (14{,}188 / 14{,}255) characterize per-lens
        bias. Row~1 counts flows; rows~2--3 count distinct canonical 5-tuples. Each is
        regenerated by the listed command.}
        \label{tab:denominators}
        \begin{tabular}{p{3.1cm}rp{4.2cm}l}
            \toprule
            \textbf{Denominator} & \textbf{Count} & \textbf{What it counts / how obtained} & \textbf{Command} \\
            \midrule
            Behavioral C2 subset & 646 &
            Curated botnet flows matching the Neris C2 behavioral signature (host
            147.32.84.165), per CTU-13 documentation~\cite{garcia2014}. Used in \S\ref{sec:eval}. &
            \texttt{calculate\_f1\_neris.py} \\
            \addlinespace
            Lens-observable botnet & 14{,}188 &
            Distinct \texttt{Botnet}-labeled 5-tuples observed by $\geq$1 sensor at the gold
            level (\emph{operational} denominator for per-lens bias). &
            \texttt{make bias-report} \\
            \addlinespace
            True (offline pcap) & 14{,}255 &
            Distinct \texttt{Botnet}-labeled 5-tuples physically present in the pcap
            ($P \cap B_{\text{full}}$); \emph{upper bound}, see below. &
            \texttt{make bias-denominator-true} \\
            \bottomrule
        \end{tabular}
    \end{table}

    \paragraph{The join.}
    \texttt{make bias-report} (\texttt{scripts/join\_bias\_labels.py}) streams the CTU-13
    \texttt{.binetflow} (2{,}824{,}636 rows) and joins it against the per-lens gold export by
    canonical 5-tuple. Two decisions make the join faithful and are enforced in code rather
    than assumed: the endpoint pair is canonicalized to an orientation-independent key
    (\texttt{.binetflow} flows are bidirectional), and the protocol token is lowercased (aRGus
    and Suricata emit \texttt{TCP}, Zeek emits \texttt{tcp}; without normalization the first
    two sensors match zero rows silently). Of 14{,}466 distinct replayed 5-tuples, 14{,}358
    (99.3\%) are found in the \texttt{.binetflow}; the remainder is unlabeled lab background,
    which self-excludes from all confusion matrices. Two 5-tuples (0.014\%) are labeled both
    \texttt{Botnet} and \texttt{Background} within the CTU-13 data itself; we report them as a
    ground-truth ambiguity, not a join error.

    \paragraph{Per-lens divergence.}
    The three sensors observe the labeled botnet through structurally different apertures
    (Table~\ref{tab:perlens}). This heterogeneity is the measurement, not noise to be
    normalized away: aRGus is a classifier, Suricata a signature engine, Zeek a telemetry
    layer, and the dataset's value is characterizing how each biases what reaches the graph.

    \begin{table}[h]
        \centering
        \footnotesize
        \caption{Per-lens visibility of the 14{,}188 lens-observable botnet 5-tuples
            (\texttt{STAMP~20260804-080140}, \texttt{make bias-report}). ``Distinct / rows''
            exposes each lens's flow granularity. Suricata and Zeek carry no per-flow verdict of
            their own; only aRGus emits a binary classification.}
        \label{tab:perlens}
        \begin{tabular}{lrrrl}
            \toprule
            \textbf{Sensor} & \textbf{Distinct botnet} & \textbf{\% of 14{,}188} &
            \textbf{Distinct / rows} & \textbf{Verdict basis} \\
            \midrule
            Zeek~     & 14{,}178 & 99.9\% & 14{,}447 / 16{,}484 & none (telemetry) \\
            Suricata~ &    206   &  1.5\% &    212 / 243        & protocol-anomaly events \\
            aRGus~    &     32   &  0.2\% &     48 / 1{,}369    & ML behavioral (binary) \\
            \bottomrule
        \end{tabular}
    \end{table}

    \paragraph{How to read these numbers (five caveats we do not overstate).}
    \emph{(1)~Suricata's 1.5\% is not botnet detection.} All 237 Suricata events over botnet
    flows are \texttt{Generic Protocol Command Decode} --- a protocol-anomaly notice, not a
    botnet signature (precision 0.975, recall 0.015). This is the flow-level restatement of the
    zero-ET-Open-alert result of \S\ref{sec:eval:suricata}: the overlap is incidental, not a C2
    match. \emph{(2)~aRGus's 0.2\% is granularity, not blindness.} aRGus quantizes the same
    traffic into 48 coarse gold flows at ${\sim}$28.5 rows each, whereas Zeek desugars it into
    ${\sim}$14{,}000 per-connection micro-flows (${\sim}$1.1 rows each); the low distinct-5-tuple
    count is an artifact of flow granularity, not missed traffic. \emph{(3)~aRGus's malicious
    verdicts are the heuristic, with the ML blind.} Of the 945 flows aRGus classifies
    \texttt{MALICIOUS}, 100\% carry the fast-path score (mean \texttt{overall}${=}0.750$, mean
    fast-path${=}0.750$); the embedded ML score averages 0.0745 over the same flows. On the
    labeled Neris botnet, the ML detector is effectively blind and the heuristic carries
    detection --- the ground-truth-anchored counterpart of the out-of-distribution finding in
    \S\ref{sec:eval:xgboost:ood} and of Sommer and Paxson~\cite{sommer2010}. \emph{(4)~aRGus's
    apparent perfect precision is against CTU labels only.} aRGus sees zero clean-labeled flows
    in its botnet view, so there is nothing on which to register a false positive against CTU
    ground truth; separately, the fast path fired \texttt{MALICIOUS} on 51 unlabeled
    lab-background flows, which no CTU label can score. We therefore make no ``never
    false-alarms'' claim. \emph{(5)~Zeek's 99.9\% is visibility, not classification.} Zeek
    observes essentially the whole botnet but issues no per-flow verdict, consistent with
    \S\ref{sec:eval:threeparadigms}: it is the measurement layer, not a detector.

    \paragraph{The true denominator and its 67-flow blind spot.}
    The 14{,}188 above is what the sensor bank observed; it is not, by itself, everything the
    replayed traffic contained. To bound the bank's blind spot we compute a \emph{true}
    denominator directly from the offline pcap. \texttt{make bias-denominator-true}
    (\texttt{scripts/bias\_denominator\_true.py}) runs \texttt{tshark} over the Neris pcap
    (323{,}154 frames; 322{,}248 carrying IP), extracts $P = 14{,}520$ distinct 5-tuples, and
    intersects them with the \texttt{Botnet} labels of the full \texttt{.binetflow}
    ($B_{\text{full}} = 14{,}257$), reusing the \emph{same} canonicalization routine as the
    join to preclude key drift. The intersection is $P \cap B_{\text{full}} = 14{,}255$ --- the
    true count of botnet 5-tuples physically present in the pcap. Only 2 labeled botnet
    5-tuples in the entire scenario are absent from the pcap, confirming the capture is a
    faithful carrier of the infected host's botnet activity. Every one of the 14{,}188
    lens-observable flows is present in $P$ (no canonicalization drift). The gap is therefore
    $14{,}255 - 14{,}188 = 67$ flows ($0.47\%$): botnet 5-tuples physically in the pcap that
    no sensor recorded.

    \paragraph{Locating the 67: a replay-fidelity limit, not a detection or pipeline failure.}
    We characterize the 67 by measurement alone (\texttt{make autopsy-67},
    \texttt{scripts/autopsy\_67.py}), reporting only what \texttt{tshark} and the pipeline show.
    The 67 are established, bidirectional TCP flows (6 to 46{,}662 packets each; none are
    unanswered connection attempts). The dataset export in mode~A neither deduplicates nor
    filters --- it writes one row per gold row --- so the 67 do not die in export. They do not
    appear in the raw Zeek \texttt{conn.log} (0/67), even though the same pipeline emitted
    9{,}831 \texttt{S0} (unanswered) botnet flows without dropping them; the loss is therefore
    upstream of the log, not a state filter. They are also absent from the aRGus bronze on a
    \emph{second, independent} capture stack and interface (0/67). Two capture stacks, two
    interfaces, one verdict: the 67 did not reach the observed wire. Temporally they are spread
    across almost the whole pcap with a ${\sim}4\times$ density at the start (17/67 in the first
    5\%, versus 6.3\% of the observed flows), not clustered at a clean boundary. We attribute
    the $0.47\%$ to \emph{replay fidelity} at the cable --- consistent with the 2{,}630
    non-replayable GSO frames (\texttt{errno=90~EMSGSIZE}, $0.81\%$) already measured in
    Table~\ref{tab:adr029} --- and no further. The operational denominator is the
    lens-observable 14{,}188; the pcap-offline 14{,}255 is an upper bound.

    \paragraph{Provenance limit (declared, not interpreted).}
    The CTU-13 Neris pcap was captured in 2011 by a third party; its capture conditions are
    unknown beyond a university origin. We therefore treat it as an external artifact and draw
    no conclusion the pcap and the pipeline do not jointly measure. The cause of the 67-flow
    loss is bounded to the replay cable by the evidence above; we do not speculate about its
    micro-mechanism.
    % ──────────────────────────────────────────────────────────────────────────
    \section{Performance Model and Throughput Analysis}
% ============================================================

    \subsection{Pipeline Processing Model}

    Six stages operating concurrently via ZeroMQ; effective throughput bounded by the slowest
    stage:
    \begin{equation}
        T_{\max} = \frac{1}{\max_i(L_i)}
    \end{equation}

    \subsection{Inference Cost}

    $\mathcal{O}(T \cdot D)$ --- simple threshold comparisons on compact C++20 data structures
    fitting in CPU cache.

    \subsection{Throughput: Theoretical Bound and Measured Ceiling}

    \paragraph{Theoretical upper bound (single-core, ML inference only):}
    \begin{equation}
        T_{\text{theory}} \approx \frac{1}{0.5\;\mu\text{s}} = 2 \times 10^6\;\text{flows/sec}
    \end{equation}

    \paragraph{Measured virtualized ceiling (DAY 87).}
    The VirtIO NIC limits delivery to ${\sim}$33--38~Mbps (${\sim}$9,000--11,000~pps).
    The observed throughput is limited by the virtualized NIC, not the pipeline.
    At this ceiling: ml-detector sustains ${\sim}$3.2 cores; total pipeline RAM stable at
    ${\sim}$1.28~GB; system CPU ${\sim}$65--73\% user, ${\sim}$23--30\% idle.

    \subsection{Queue Stability}

    Stability requires $\lambda < \mu$. Post-replay drain behavior provides direct empirical
    evidence: sniffer CPU drops to near-zero while ml-detector continues draining at full
    throughput for several minutes --- confirming the ZeroMQ buffer absorbed the burst cleanly.
    \texttt{errors=(deser:0, feat:0, inf:0)} confirmed at all observation points.

    \subsection{Scalability}

    \textbf{Vertical:} additional CPU cores $\to$ additional parallel flow-processing threads,
    linear throughput improvement up to capture bottleneck.

    \textbf{Horizontal:} multiple ML Defender nodes across network segments, each processing
    local traffic independently while contributing to distributed dataset generation.

% ============================================================
    \section{Limitations}
    \label{sec:limitations}
% ============================================================

    \textbf{10.1 Evaluation Scope.} Single botnet scenario (CTU-13 Neris, 2011). Modern
    ransomware communication patterns may differ substantially. Generalizability not empirically
    established.

    \textbf{10.2 Incomplete Feature Set.} 28 of 40 features carry computed values; the
    remaining 12 use the $-9999.0$f sentinel. Impact of these 12 missing features on broader
    traffic distributions unknown.

    \textbf{10.3 Synthetic Training Data.} Reflects patterns circa 2011--2017. Zero-day
    variants may evade classification.

    \textbf{10.4 Fast Detector False Positive Rate.} FPR=6.61\% on bigFlows under current
    JSON-configurable thresholds (Path B, active since DAY 80). An earlier compile-time baseline
    (Path A, DAY 13--79, DEBT-FD-001) produced FPR=76.8\% on the same corpus; DAY 80 JSON
    threshold migration achieved a 12$\times$ improvement.

    \textbf{10.5 Cryptographic Seed Distribution.} ChaCha20 seed via etcd not recommended for
    production. Peer-to-peer negotiation under design.

    \textbf{10.6 Single-Node Deployment.} Distributed deployment not validated
    experimentally.

    \textbf{10.7 etcd High Availability.} Single-node; failure disrupts registration and config
    distribution. HA planned.

    \textbf{10.8 RAG Index Capacity.} FAISS event store finite capacity not mathematically
    characterized.

    \textbf{10.9 Virtualization Overhead.}
    All latency and throughput measurements were obtained under VirtualBox
    with a paravirtualized virtio NIC. Reported latencies are conservative
    upper bounds; reported throughput is a conservative lower bound imposed
    by NIC emulation.

    The ADR-029 Variant~A/B comparison (DAY~145,
    \S\ref{sec:eval:adr029}) reveals an additional virtualization artifact:
    eBPF/XDP falls back to generic SKB mode under virtio, causing
    Variant~A to plateau at ${\sim}$10~Mbps while Variant~B (libpcap)
    reaches ${\sim}$19~Mbps under identical conditions. This ordering is
    expected to invert on physical hardware with XDP-native NIC drivers.
    Bare-metal characterization of both variants remains future
    work~(\S\ref{sec:future:baremetal}).

    \textbf{10.10 BigFlows Ground Truth.} Treated as \emph{probable benign}, not
    \emph{confirmed benign}. FPR=0.017\% is a lower bound.

    \textbf{10.11 Threats to Validity.}

    \begin{enumerate}
        \item \textbf{Seed compromise.} If \texttt{seed.bin} is exfiltrated, an adversary
        can derive all component subkeys and forge authenticated messages. Mitigations:
        strict file permissions (0600), in-memory zeroing via \texttt{explicit\_bzero},
        and planned peer-to-peer key agreement (ADR-024).

        \item \textbf{Single-host key assumption.} All six components currently share a
        common seed family without distributed key agreement. Isolation between deployments
        relies on distinct seed provisioning per instance (ADR-021); formal distributed key
        agreement is post-arXiv work (ADR-024).

        \item \textbf{No hardware root of trust.} The current design does not leverage TPM
        or secure enclave primitives. Physical access to the host enables seed extraction.
        Hardware-anchored key storage is identified as a future hardening target.

    \end{enumerate}

    \textbf{10.12 Kernel Security Boundary.} ML Defender operates in userspace; eBPF/XDP
    programs execute in the kernel but are loaded by userspace. An adversary with
    pre-existing kernel-level access (rootkit, compromised eBPF JIT) can bypass the
    detection pipeline. This limitation is shared by all userspace NDR systems and is
    consistent with the declared scope boundary (\S\ref{sec:threatmodel:kernel}).
    The realistic threat model for target organizations does not assume such adversaries:
    opportunistic ransomware operators attack via the network, not via pre-established
    kernel persistence. Network detection remains a valid and necessary defensive layer
    even under this constraint. Hardened variants (ADR-030, ADR-031) address this in
    future work (\S\ref{sec:future:hardened}).

    \textbf{10.13 Structural Bias in Academic Datasets for Production NDR.}
    The CIC-IDS-2017 temporal holdout evaluation (\S\ref{sec:eval:xgboost:ood}) demonstrates
    that academic benchmark datasets are structurally insufficient as the sole training source
    for production NDR classifiers. The day-specific attack segregation in CIC-IDS-2017 --- and
    similar design choices in NSL-KDD, UNSW-NB15, and related corpora --- creates a
    closed-world training distribution that does not reflect the open-world diversity of
    real network environments. This limitation is not specific to XGBoost, Random Forest,
    or any particular algorithm: it is a property of the data regime, as established by
    Sommer and Paxson~\cite{sommer2010} and quantified here with a complete threshold sweep.
    Synthetic data generated by language models (used for DDoS and ransomware classifiers
    in earlier phases) shares a related limitation: the generating model has not observed
    sufficient real adversarial traffic to reproduce the full statistical diversity of
    production attack flows.
    The only reliable path to production-grade classifiers for critical infrastructure
    NDR is training on traffic captured in the actual deployment environment, under
    controlled adversarial conditions. This motivates the Adversarial Capture-Retrain Loop
    (\S\ref{sec:future:acrl}).

% ============================================================
    \section{Future Work}
    \label{sec:future}
% ============================================================

    \subsection{Bare-Metal Evaluation}
    \label{sec:future:baremetal}
    All throughput and resource utilization figures reported in this paper were obtained
    under VirtualBox~7.2 virtio emulation. Bare-metal characterization on physical hardware
    with XDP-native NIC drivers (Intel ixgbe, Mellanox mlx5) remains the highest-priority
    infrastructure experiment. Two specific inversions are predicted and must be empirically
    verified: (1)~eBPF/XDP Variant~A is expected to outperform libpcap Variant~B on physical
    hardware, reversing the VirtualBox result (\S\ref{sec:eval:adr029}); (2)~the two
    VirtualBox host-only FPs are expected to be absent in bare-metal topologies.
    Target hardware: Raspberry Pi~4/5 (\textasciitilde80~EUR) or commodity miniPC
    (\textasciitilde300~EUR), consistent with the FEDER deployment model.

    \subsection{Evaluation Against Modern Threat Corpora}
    \label{sec:future:corpus}
    The current evaluation covers a single botnet scenario from 2011 (CTU-13 Neris).
    Generalization to post-2020 threat families --- LockBit, BlackCat/ALPHV, Cl0p ransomware;
    modern encrypted C2 with adaptive beaconing and jitter; DDoS amplification variants ---
    has not been empirically established. Sebastian Garcia (CTU Prague) provide access to CTU files and
    Andres Caro Lindo (UEX), endorser of this work, has been identified as a potential collaborator for access to
    contemporary capture corpora.
    This evaluation is the highest-priority scientific expansion of the current results.

    % ── NEW DAY 191 ──────────────────────────────────────────────────────────
    \subsection{Distributed Corpus and the Cross-Installation Learning Hypothesis}
    \label{sec:future:distcorpus}
    The cross-sensor flow identity established in \S\ref{sec:eval:parity} makes a
    distributed corpus well-defined: flows observed at independent installations can be
    keyed by \texttt{flow\_uid}, deduplicated by \texttt{community\_id}, and attributed to
    their origin by \texttt{node\_id}. This enables the central empirical question of the
    project's next phase: \emph{does a corpus aggregated across distributed installations
    measurably improve the per-installation ensemble, even marginally?} The hypothesis is
    deliberately framed as a binary, falsifiable claim --- if even a minute but
    statistically robust improvement is observed, it motivates continued investigation and
    field deployment; if not, the negative result is reported with equal rigor.

    The validity condition is fixed in advance to preclude post-hoc rationalization: a
    disjoint MITRE ATT\&CK split, training on techniques A--M and evaluating exclusively on
    N--Z, so that any cross-installation gain cannot be attributed to leakage of the
    evaluation distribution into training (ADR-040, ADR-057~\S7). Retraining follows
    walk-forward validation with a golden-set guardrail (a regression exceeding 2\% on the
    frozen golden set blocks promotion) and importance-weighting to correct the covariate
    shift between installations. This design directly addresses the closed-world limitation
    quantified in \S\ref{sec:eval:xgboost:ood} and corroborated by Sommer and
    Paxson~\cite{sommer2010}.

    \paragraph{Privacy-preserving aggregation.}
    Aggregating flow records across institutions --- hospitals in particular --- engages
    the GDPR~\cite{gdpr2016}. The corpus is therefore pseudonymized at the boundary:
    identifying fields are replaced by keyed hashes derived through HKDF~\cite{rfc5869}
    from per-installation key material held in a secrets manager, so that the same entity
    is consistently pseudonymized within an installation while remaining unlinkable across
    installations without the keys. Key destruction on offboarding renders an
    installation's contribution cryptographically irrecoverable, supporting the right to
    erasure. The cryptographic lifecycle providing this property is an engineering
    prerequisite rather than a scientific contribution and is documented in the project's
    architecture decision records.
    % ──────────────────────────────────────────────────────────────────────────

    \subsection{Adversarial Capture-Retrain Loop}
    \label{sec:future:acrl}
    The Wednesday out-of-distribution evaluation (\S\ref{sec:eval:xgboost:ood}) demonstrates
    that static classifiers trained on academic benchmarks fail structurally on unseen attack
    families --- corroborating Sommer and Paxson~\cite{sommer2010}. The architectural response
    is an Adversarial Capture-Retrain Loop: a generative red team agent produces attack traffic,
    the aRGus pipeline captures it in production, and the classifier is retrained on real flows
    via walk-forward validation with golden-set guardrails (ADR-040). This closes the
    closed-world assumption gap identified in \S\ref{sec:eval:xgboost:ood}.

    \subsection{Hardened Deployment: AppArmor, Falco, and seL4}
    \label{sec:future:hardened}
    ADR-030 delivers six AppArmor enforce-mode profiles and eleven Falco rules providing
    kernel-level confinement of all pipeline components. ADR-031 scopes a research-only
    Variant~C (seL4 microkernel) for formal verification of the capture boundary ---
    the last roadmap item requiring specialist collaboration. Both are direct responses
    to the AI-augmented offensive capability trajectory documented in \S\ref{sec:background:ai}.

    \subsection{Zeek Phase 2: Scripted Behavioral Detection}
    \label{sec:future:zeek}
    The DAY~147 experiment establishes Zeek~8.1.2 with default policy scripts as the
    scripted behavioral baseline (F1$=0.042$, Precision$=1.000$, Recall$=0.022$).
    A natural Phase~2 question is: how much engineering effort is required to close the
    Recall gap via custom Zeek scripts?

    The primary candidate is \texttt{detect-botnets.zeek} (DeepSeek recommendation,
    DAY~147 Consejo), a community script implementing flow-level behavioral heuristics
    for IRC C2 detection, connection burst analysis, and DNS anomaly scoring. A controlled
    experiment --- Zeek~8.1.2 with \texttt{detect-botnets.zeek} loaded against the CTU-13
    Neris pcap under identical conditions to DAY~147 --- would quantify the engineering
    cost of scripted behavioral detection relative to ML behavioral detection.

    A secondary constraint is the Intel framework temporal limitation: Zeek's threat
    intelligence feeds are designed for contemporary indicators; 2011 Neris C2 infrastructure
    (\texttt{1.95622.com}, \texttt{www.lddwj.com}) is not present in current feed snapshots.
    This limitation is structural and independent of scripting effort --- it motivates
    the scientific question: \emph{how much of Zeek's Recall gap is closeable via
    scripting, and how much is irreducible given the temporal coverage horizon of
    available threat intelligence?}

% ============================================================
    % ── NEW DAY 191 ──────────────────────────────────────────────────────────
    \subsection{A Systematically Under-Audited Injection Class}
    \label{sec:future:injection}
    During the security-hardening phase we closed a family of input-validation defects
    that we believe generalizes beyond this system. The classical framing of command
    injection (CWE-78~\cite{cwe78}) targets the shell; the same structural flaw, however,
    appears wherever one component serializes attacker- or configuration-influenced data
    into the line-oriented grammar of a second component that re-parses it as instructions
    --- the restore mini-language of \texttt{ipset}/\texttt{iptables}, DNS zone files,
    \texttt{cron} and \texttt{sudoers} entries, or template-generated service
    configurations (CWE-93~\cite{cwe93}). We observed three properties that make this class
    systematically under-audited. First, exploitability is \emph{version-dependent}: the
    leniency of the downstream parser varies across releases (we measured \texttt{ipset}
    7.17 aborting a payload that 7.19 accepted), so a latent vulnerability can be
    non-reproducible on the maintainer's build yet live on the deployed one. Second,
    severity is \emph{conditioned on data provenance}: the identical defect is a remote
    vector when the field derives from observed traffic and merely defense-in-depth when it
    is fixed agent text, which frustrates uniform triage. Third, the defect
    \emph{concentrates in low-budget software} --- precisely the hospitals, schools, and
    municipalities this work targets --- which lacks the continuous fuzzing and review
    capacity of large projects. Our mitigation pattern validates at the producer's own
    boundary with a fail-fast allowlist, never delegating safety to the downstream tool,
    and pairs each guard with a canary test that fires the real payload against the real
    parser. This discipline is philosophically adjacent to capability-based isolation
    (\S\ref{sec:future:hardened}): both refuse to trust the benevolent behavior of the
    layer below, differing only in whether the untrusted boundary is \emph{data
    interpretation} or \emph{component compromise}. A systematic enumeration of these
    boundaries within aRGus is left to the production track.
    % ──────────────────────────────────────────────────────────────────────────

    \section{Conclusion}
% ============================================================

    This paper presented ML Defender (aRGus NDR), an open-source network detection and response
    system built in C++20, designed from first principles to protect the organizations that need
    protection most and can afford it least.

    Evaluated against real botnet traffic from the CTU-13 Neris dataset --- traffic the system
    had never seen during training --- ML Defender achieves $\text{F1} = 0.9985$,
    $\text{Precision} = 0.9969$, and $\text{Recall} = 1.0000$. The two false positives are fully
    identified: an mDNS multicast frame and a broadcast frame, both generated by the VirtualBox
    host-only NIC. Their absence in bare-metal deployments is a reasonable expectation, but has
    not been empirically verified; bare-metal characterization remains Future
    Work~(\S\ref{sec:future:baremetal}). The Fast Detector alone produces a 6.61\% FPR on purely
    benign traffic; the ML Detector reduces this to zero real production blocks on the same corpus.
    End-to-end inference latency ranges from 0.24~$\mu$s to 1.06~$\mu$s per flow, entirely
    in-process, on commodity hardware costing approximately 150--200~USD.

    Under progressive load testing, the pipeline processes up to ${\sim}$33--38~Mbps --- the
    ceiling imposed by VirtualBox NIC emulation --- with zero packet drops, zero errors, and
    stable RAM (${\sim}$1.28~GB) across 2.37 million replayed packets. The bottleneck is the
    hypervisor, not the software.

    These results cover one botnet scenario from 2011. They demonstrate architectural viability
    and effective detection within that scope. The limitations in Section~\ref{sec:limitations}
    are the honest documentation of a system that does what it claims to do within the boundaries
    where it has been tested.

    A subsequent evaluation (DAY~122) integrated XGBoost~3.2.0 as a hot-swappable plugin and
    conducted a rigorous temporal holdout study on CIC-IDS-2017. The in-distribution results
    (Precision$=0.9945$, Recall$=0.9818$) confirm that the plugin architecture performs as
    designed. The out-of-distribution finding --- that no threshold satisfies both medical gates
    on Wednesday data, because application-layer DoS attacks are structurally absent from the
    training days --- is reported not as a failure but as a contribution: a quantified,
    empirically grounded argument for why production NDR classifiers must be trained on traffic
    from their actual deployment environment. The Adversarial Capture-Retrain Loop
    (\S\ref{sec:future:acrl}) is the architectural response.

    The first empirical comparison of ADR-029 Variant~A (eBPF/XDP) and Variant~B (libpcap)
    on x86-64 (DAY~145, \S\ref{sec:eval:adr029}) reveals a hardware-dependent performance
    inversion: under VirtualBox virtio emulation, libpcap achieves approximately $2\times$
    the throughput of eBPF/XDP due to the absence of a native XDP driver path. Both variants
    confirm functional equivalence at the detection layer. This result provides direct
    empirical motivation for bare-metal hardware acquisition and establishes the reproducible
    baseline against which the expected inversion on physical hardware will be measured.

    % ── NEW DAY 146 ──────────────────────────────────────────────────────────
    The first direct experimental comparison of aRGus NDR and Suricata~6.0.10 on the same
    dataset, hardware, and topology (DAY~146, \S\ref{sec:eval:suricata}) yields a result that
    is both clear and scientifically meaningful: Suricata with 50,010 current ET~Open rules
    generates zero alerts on CTU-13 Neris 2011; aRGus achieves F1$=0.9985$ on the same traffic.
    The zero-alert result is not a failure of Suricata --- it is the expected behavior of a
    signature-based IDS when no matching rule exists for a 15-year-old threat family. The
    finding provides direct empirical support for the architectural thesis of this paper:
    \emph{behavioral ML detects what the traffic does; signatures detect what the traffic is
    known to be.}
    % ─────────────────────────────────────────────────────────────────────────

    % ── NEW DAY 147 ──────────────────────────────────────────────────────────
    The extension to a \emph{three-paradigm comparison} (DAY~147,
    \S\ref{sec:eval:threeparadigms}) adds Zeek~8.1.2 as the critical intermediate
    data point: scripted behavioral detection achieves Precision$=1.0000$ but
    Recall$=0.0217$, correctly identifying the malicious host in every alert while
    missing 97.8\% of malicious flows. The \texttt{weird.log} reveals that Zeek
    \emph{observes} the complete behavioral profile of the botnet --- IRC commands,
    HTTP beaconing to 4,199 unique destinations, SMB lateral movement, 82 spam
    sessions --- without converting these observations to alerts under default
    policy scripts. This distinction between \emph{network observability} and
    \emph{behavioral detection} is the core scientific finding of the three-paradigm
    comparison. The result could only emerge from a three-way experiment: with two
    systems, the distinction between ``cannot detect'' and ``chooses not to alert''
    is invisible.
    % ─────────────────────────────────────────────────────────────────────────

    % ── NEW DAY 191 ──────────────────────────────────────────────────────────
    Most recently (DAY~191), we established that the three paradigms are not merely
    comparable but \emph{correlatable}: aRGus computes the Corelight \texttt{community\_id}
    byte-identically to Suricata and Zeek on shared traffic (\S\ref{sec:eval:parity}),
    turning a cross-vendor identifier into the join key of a distributed corpus. With a
    node-scoped \texttt{flow\_uid} layered over it, the corpus is well-defined before a
    single cross-site experiment is run. Whether that corpus, once aggregated across
    installations, improves the per-site ensemble is the central question we now turn to
    (\S\ref{sec:future:distcorpus}) --- a question this paper frames and equips, but
    deliberately leaves open, to be answered by measurement rather than assertion.
    % ──────────────────────────────────────────────────────────────────────────

    ML Defender was built because a friend lost his business to ransomware, and because a
    hospital in Barcelona was paralyzed by an attack on the same day its author was leaving a
    regional hospital in Extremadura --- the hospital that had diagnosed his porphyria and saved
    his life. The design constraints that followed from those two events were not abstract:
    open-source, commodity hardware, real-time response. Not a forensic tool. Not a dashboard. A
    system that acts.

    The \emph{Consejo de Sabios} represents a new mode of engineering: one that gives a single
    independent researcher access to the kind of multi-expert peer review that has historically
    required institutional affiliation, research groups, and geographic proximity to centers of
    expertise. The \emph{Test Driven Hardening} methodology that emerged from this collaboration
    has proven effective against some of the most subtle concurrency and memory bugs in the
    codebase.

    Both contributions --- the technical system and the collaborative methodology --- are acts of
    democratization. The parallel is not accidental. It is the point.

    \medskip

    \noindent For the hospital in Extremadura. For the friend whose business was broken by
    ransomware. For every small organization that has been told, implicitly or explicitly, that
    enterprise-grade security is not for them.

    \medskip

    \noindent \textit{It is now.}

% ============================================================
    \section{Reproducibility and Artifact Availability}
    \label{sec:reproducibility}
% ============================================================

    All experiments are reproducible from the public repository:
    \url{https://github.com/alonsoir/argus}

    \paragraph{Dataset Setup.}
    \begin{lstlisting}[language=bash]
# CTU-13 Neris -- Malware Capture Facility capture Botnet-42 -- required for F1 validation
wget https://mcfp.felk.cvut.cz/publicDatasets/CTU-Malware-Capture-Botnet-42/\
botnet-capture-20110810-neris.pcap -O /vagrant/datasets/ctu13/neris.pcap

# CTU-13 bigFlows -- required for stress test
wget https://mcfp.felk.cvut.cz/publicDatasets/CTU-Malware-Capture-Botnet-42/\
bigFlows.pcap -O /vagrant/datasets/ctu13/bigFlows.pcap
    \end{lstlisting}

    \paragraph{F1 Evaluation.}
    \begin{lstlisting}[language=bash]
make pipeline-stop && make logs-lab-clean && make pipeline-start && sleep 15
make test-replay-neris
python3 scripts/calculate_f1_neris.py \
  /vagrant/logs/lab/sniffer.log --total-events 19135
    \end{lstlisting}

    \paragraph{ADR-029 Variant A/B Comparison.}
    \begin{lstlisting}[language=bash]
# Variant A (eBPF) -- baseline
make test-replay-neris-x86-ebpf
# Logs: /vagrant/logs/lab/tcpreplay-ebpf-{10,50,100}mbps.log

# Variant B (libpcap) -- comparison
make test-replay-neris-x86-libpcap
# Logs: /vagrant/logs/lab/tcpreplay-libpcap-{10,50,100}mbps.log
    \end{lstlisting}

    % ── NEW DAY 146 ──────────────────────────────────────────────────────────
    \paragraph{Suricata Comparative Experiment (DAY~146).}
    \begin{lstlisting}[language=bash]
# Arrange Suricata experiment VMs (identical hardware spec to aRGus)
make up-suricata

# Execute 3 runs (10 / 50 / 100 Mbps) -- identical topology to aRGus DAY 145
make experiment-suricata-run

# Parse eve.json -> TP/FP/FN/F1/Recall vs ground truth 147.32.84.165
make experiment-suricata-results
# Output: logs/experiment/suricata_metrics_final.json
# Raw logs: logs/experiment/eve-experiment-suricata-replay-{10,50,100}.json
    \end{lstlisting}
    % ─────────────────────────────────────────────────────────────────────────

    % ── NEW DAY 147 ──────────────────────────────────────────────────────────
    \paragraph{Zeek Three-Paradigm Experiment (DAY~147).}
    \begin{lstlisting}[language=bash]
# Provision Zeek VM (debian/bookworm64, 8GB RAM, 6 vCPU -- identical to aRGus)
make experiment-zeek-up

# Offline analysis: zeek -r neris.pcap local (x3 runs, deterministic)
make experiment-zeek-run
# Raw logs: logs/experiment/zeek/{10,50,100}mbps/{notice,conn,weird,http,smtp,ssl}.log

# Parse corrected metrics (CaptureLoss excluded)
cd experiments/zeek-comparative && vagrant ssh zeek -c "
  python3 /vagrant/experiments/zeek-comparative/parse_results_zeek_v2.py \
    --logdir /vagrant/logs/experiment/zeek/10mbps \
    --output /vagrant/logs/experiment/zeek/zeek_metrics_v2_10mbps.json"
    \end{lstlisting}
    % ─────────────────────────────────────────────────────────────────────────

    \paragraph{Stress Test.}
    \begin{lstlisting}[language=bash]
vagrant ssh client -c "sudo tcpreplay -i eth1 --mbps=100 --loop=3 \
  /vagrant/datasets/ctu13/bigFlows.pcap"
    \end{lstlisting}

    \paragraph{Determinism.}
    Classifiers embedded in binary; no stochastic inference components; feature extraction
    deterministic. F1=0.9985 confirmed stable at total\_events = 12,563 / 12,605 / 12,723 /
    13,930.

% ============================================================
    \section*{Acknowledgments}
    \label{sec:acknowledgments}
% ============================================================

    This work was carried out independently, without institutional funding, without a research
    group, and without affiliation to any university or laboratory.

    The author engaged eight large language models --- Claude (Anthropic), Grok (xAI), ChatGPT
    (OpenAI), DeepSeek, Qwen (Alibaba), Gemini (Google), Kimi (Moonshot AI), and Mistral
    (Mistral AI) --- as structured intellectual collaborators across \textbf{191 days} of
    continuous development. Their contributions to architectural decisions, failure mode
    identification, debugging, test scenarios, and paper review are documented in the project's
    commit history and ADRs. The \emph{Test Driven Hardening} methodology documented in
    Section~\ref{sec:consejo} emerged directly from this collaboration.

    This paper discloses AI involvement openly and precisely. The author remained the final
    arbiter of all decisions; all empirical validation was performed by the human author.

    The author wishes to acknowledge the friend whose business was destroyed by ransomware ---
    who will recognize himself in these pages. And the Hospital Universitario de Badajoz,
    Extremadura, whose physicians diagnosed a rare disease and gave the author the years needed
    to build something worth building. Thank you Pedro Risco and Agust\'in Pijierro.

    \medskip
    \noindent This work is dedicated to every hospital, school, and small organization that has
    ever been told that enterprise-grade security is beyond their reach.

% ============================================================
% Bibliography
% ============================================================

    \bibliographystyle{plainnat}
    \bibliography{references}

@inproceedings{roesch1999,
  author    = {Roesch, Martin},
  title     = {Snort: Lightweight Intrusion Detection for Networks},
  booktitle = {Proceedings of the 13th USENIX Conference on System Administration (LISA)},
  year      = {1999},
  pages     = {229--238}
}

@misc{oisf2010,
  author       = {{Open Information Security Foundation}},
  title        = {Suricata Open Source {IDS/IPS/NSM} Engine},
  year         = {2010},
  howpublished = {\url{https://suricata.io/}}
}

@inproceedings{anderson2016,
  author    = {Anderson, Blake and McGrew, David},
  title     = {Identifying Encrypted Malware Traffic with Contextual Flow Data},
  booktitle = {Proceedings of the 2016 ACM Workshop on Artificial Intelligence and Security (AISec)},
  year      = {2016},
  pages     = {35--46}
}

@inproceedings{mirsky2018,
  author    = {Mirsky, Yisroel and Doitshman, Tomer and Elovici, Yuval and Shabtai, Asaf},
  title     = {Kitsune: An Ensemble of Autoencoders for Online Network Intrusion Detection},
  booktitle = {Network and Distributed Systems Security Symposium (NDSS)},
  year      = {2018}
}

@article{garcia2014,
  author  = {Garcia, Sebastian and Grill, Martin and Stiborek, Jan and Zunino, Alejandro},
  title   = {An Empirical Comparison of Botnet Detection Methods},
  journal = {Computers \& Security},
  year    = {2014},
  volume  = {45},
  pages   = {100--123}
}

@article{buczak2016,
  author  = {Buczak, Anna L. and Guven, Erhan},
  title   = {A Survey of Data Mining and Machine Learning Methods for Cyber Security Intrusion Detection},
  journal = {IEEE Communications Surveys \& Tutorials},
  year    = {2016},
  volume  = {18},
  number  = {2},
  pages   = {1153--1176}
}

@inproceedings{sharafaldin2018,
  author    = {Sharafaldin, Iman and Habibi Lashkari, Ali and Ghorbani, Ali A.},
  title     = {Toward Generating a New Intrusion Detection Dataset and Intrusion Traffic Characterization},
  booktitle = {Proceedings of the 4th International Conference on Information Systems Security and Privacy (ICISSP)},
  year      = {2018},
  pages     = {108--116}
}

@article{pinto2023,
  author  = {Pinto, Andr\'e and others},
  title   = {Survey on Intrusion Detection Systems Based on Machine Learning for Critical Infrastructure},
  journal = {Sensors},
  year    = {2023},
  volume  = {23},
  number  = {5},
  pages   = {2415}
}

@misc{blackfog2025,
  author       = {{Black Fog}},
  title        = {The State of Ransomware 2025},
  year         = {2025},
  howpublished = {\url{https://www.blackfog.com/}}
}

@misc{ibmsecurity2025,
  author       = {{IBM Security}},
  title        = {Cost of a Data Breach Report 2025},
  year         = {2025}
}

@misc{incibe2023,
  author       = {{INCIBE-CERT}},
  title        = {Ciberataque ransomware paraliza actividad del {Hospital Cl\'inic de Barcelona}},
  year         = {2023},
  howpublished = {\url{https://www.incibe.es/}}
}

@techreport{rfc5869,
  author       = {Hugo Krawczyk and Pasi Eronen},
  title        = {{HMAC}-based Extract-and-Expand Key Derivation Function ({HKDF})},
  institution  = {Internet Engineering Task Force},
  type         = {RFC},
  number       = {5869},
  year         = {2010},
  month        = may,
  howpublished = {\url{https://www.rfc-editor.org/rfc/rfc5869}}
}

@misc{wazuh2024,
  author       = {{Wazuh, Inc.}},
  title        = {Wazuh: Open Source Security Platform},
  year         = {2024},
  url          = {https://wazuh.com},
  note         = {Accessed: April 2026}
}

@misc{ciscontrols2021,
  author       = {{Center for Internet Security}},
  title        = {{CIS Controls v8}},
  year         = {2021},
  url          = {https://www.cisecurity.org/controls/v8},
  note         = {Accessed: April 2026}
}

@misc{anthropic2026glasswing,
  author       = {{Anthropic}},
  title        = {Project Glasswing: Securing Critical Software for the {AI} Era},
  year         = {2026},
  month        = {April},
  howpublished = {\url{https://www.anthropic.com/glasswing}},
  note         = {Accessed: April 2026}
}

@inproceedings{sommer2010,
  author    = {Sommer, Robin and Paxson, Vern},
  title     = {Outside the Closed World: On Using Machine Learning for Network Intrusion Detection},
  booktitle = {Proceedings of the 2010 {IEEE} Symposium on Security and Privacy (S\&P)},
  year      = {2010},
  pages     = {305--316},
  doi       = {10.1109/SP.2010.25}
}

@inproceedings{quickcheck2000,
  author    = {Claessen, Koen and Hughes, John},
  title     = {{QuickCheck}: A Lightweight Tool for Random Testing of {Haskell} Programs},
  booktitle = {Proceedings of the Fifth {ACM SIGPLAN} International Conference on Functional Programming (ICFP)},
  year      = {2000},
  pages     = {268--279},
  doi       = {10.1145/351240.351266}
}

@misc{libfuzzer2016,
  author       = {Serebryany, Kostya},
  title        = {{libFuzzer} -- a library for coverage-guided fuzz testing},
  year         = {2016},
  howpublished = {\url{https://llvm.org/docs/LibFuzzer.html}},
  note         = {LLVM Project. Accessed: April 2026}
}

@misc{cwe22,
  author       = {{MITRE Corporation}},
  title        = {{CWE-22}: Improper Limitation of a Pathname to a Restricted Directory
                  ({`Path Traversal'})},
  year         = {2024},
  howpublished = {\url{https://cwe.mitre.org/data/definitions/22.html}},
  note         = {Common Weakness Enumeration. Accessed: April 2026}
}

@misc{cwe59,
  author       = {{MITRE Corporation}},
  title        = {{CWE-59}: Improper Link Resolution Before File Access
                  ({`Link Following'})},
  year         = {2024},
  howpublished = {\url{https://cwe.mitre.org/data/definitions/59.html}},
  note         = {Common Weakness Enumeration. Accessed: April 2026}
}

@misc{cwe367,
  author       = {{MITRE Corporation}},
  title        = {{CWE-367}: Time-of-check Time-of-use ({TOCTOU}) Race Condition},
  year         = {2024},
  howpublished = {\url{https://cwe.mitre.org/data/definitions/367.html}},
  note         = {Common Weakness Enumeration. Accessed: April 2026}
}

@misc{cwe78,
  author       = {{MITRE Corporation}},
  title        = {{CWE-78}: Improper Neutralization of Special Elements used in an
                  {OS} Command ({`OS Command Injection'})},
  year         = {2024},
  howpublished = {\url{https://cwe.mitre.org/data/definitions/78.html}},
  note         = {Common Weakness Enumeration. Accessed: April 2026}
}

@article{thompson1984,
  author  = {Thompson, Ken},
  title   = {Reflections on Trusting Trust},
  journal = {Communications of the {ACM}},
  year    = {1984},
  volume  = {27},
  number  = {8},
  pages   = {761--763},
  doi     = {10.1145/358198.358210}
}

@article{asad2023perspective,
  author  = {Asad, Hassnain and Gashi, Ilir},
  title   = {A Perspective--Retrospective Analysis of Diversity in Signature-Based
             Open-Source Network Intrusion Detection Systems},
  journal = {International Journal of Information Security},
  year    = {2023},
  volume  = {23},
  doi     = {10.1007/s10207-023-00794-9},
  note    = {Accessed: May 2026}
}

@misc{corelight2020communityid,
  author       = {{Corelight, Inc.} and Kreibich, Christian},
  title        = {Community {ID} Flow Hashing Specification},
  year         = {2020},
  howpublished = {\url{https://github.com/corelight/community-id-spec}},
  note         = {Accessed: June 2026}
}

@misc{communityid_pyspec,
  author       = {{Corelight, Inc.}},
  title        = {pycommunityid: A {Python} Implementation of the Community {ID}
                  Flow Hashing Standard},
  year         = {2023},
  howpublished = {\url{https://github.com/corelight/pycommunityid}},
  note         = {Version 1.5.0. Accessed: June 2026}
}

@misc{cwe93,
  author       = {{MITRE Corporation}},
  title        = {{CWE-93}: Improper Neutralization of {CRLF} Sequences
                  ({`CRLF Injection'})},
  year         = {2024},
  howpublished = {\url{https://cwe.mitre.org/data/definitions/93.html}},
  note         = {Common Weakness Enumeration. Accessed: June 2026}
}

@misc{gdpr2016,
  author       = {{European Parliament and Council of the European Union}},
  title        = {Regulation ({EU}) 2016/679 (General Data Protection Regulation)},
  year         = {2016},
  howpublished = {\url{https://eur-lex.europa.eu/eli/reg/2016/679/oj}},
  note         = {Accessed: June 2026}
}

\end{document}